\documentclass[aps,prb,reprint,twocolumn,showpacs,superscriptaddress]{revtex4-2}

\usepackage{amsmath}
\usepackage{graphicx}
\usepackage{lmodern}
\usepackage{amsmath}
\usepackage{color}
\usepackage{amssymb}
\usepackage{bm}
\usepackage{braket}
\usepackage{mathtools}

\newcommand{\br}{\bm{r}}
\newcommand{\bk}{\bm{k}}
\newcommand{\bp}{\bm{p}}
\newcommand{\bq}{\bm{q}}

\usepackage[dvipsnames]{xcolor}
\usepackage[colorlinks=true]{hyperref}
\hypersetup{
    colorlinks=true,
    linkcolor=blue,
    citecolor=blue,
    urlcolor=blue
} 

\newcommand{\jjy}[1]{{\color{black}{#1}}}

\makeatletter
\def\maketitle{
\@author@finish
\title@column\titleblock@produce
\suppressfloats[t]}
\makeatother

\begin{document}
\title{shot noise in a phenomenological model of a marginal Fermi liquid}

\author{Yi-Ming Wu}
\affiliation{Stanford Institute for Theoretical Physics, Stanford University, Stanford, California 94305, USA}
\author{Josephine J. Yu}
\affiliation{Stanford Institute for Theoretical Physics, Stanford University, Stanford, California 94305, USA}
\author{S. Raghu}
\affiliation{Stanford Institute for Theoretical Physics, Stanford University, Stanford, California 94305, USA}

\date{\today}

\begin{abstract}
The strange metal is a mysterious non-Fermi liquid which shows linear-in-$T$ resistivity behavior at finite temperatures, and, as found in recent experiment, vanishingly small shot noise in the linear-in-$T$ regime. 
Here, we investigate the shot noise of a strange metal based on a phenomenological model of marginal Fermi liquid (MFL), where fermions couple to some collective boson mode, leading to $T$-linear scattering rate at finite $T$. It is found that in the diffusive regime where the MFL scattering length is small compared to the system size, the shot noise vanishes, and the thermal noise becomes a temperature- and voltage-independent constant. Introducing additional impurity scattering increases the shot noise, and is probably consistent with the current experiment.   
\end{abstract}

\maketitle

\date{\today}

\section{Introduction} 
\label{sec:introduction}

Strange metals are highly entangled electronic systems that display unusual transport behaviors due to the lack of coherent quasiparticles. Perhaps the most mysterious and ubiquitous signature of a strange metal is the linear-in-$T$ resistivity that persists over a wide temperature range, as observed in distinct materials including 
cuprates\cite{PhysRevLett.69.2975,PhysRevLett.59.1337,Keimer2015,10.21468/SciPostPhys.6.5.061,RevModPhys.92.031001,Yuan2022}, heavy-fermion materials\cite{Hilbert996,PhysRevLett.72.3262,Park2006,PhysRevLett.85.626,PhysRevLett.89.056402}, iron-based superconductors\cite{PhysRevB.81.184519,Hayes2016,annurev1,Jiang2023}, and even twisted bilayer graphene\cite{PhysRevLett.124.076801}. 
Strange metal behavior appears also to be intimately related to high-$T_c$ superconductivity (especially in cuprates), as the strange metal phase usually develops in the normal state regime right above the underlying superconducting dome. 
The recent experimental measurement of shot noise of the heavy fermion quantum critical metal YbRh$_2$Si$_2$ (YRS)  \cite{doi:10.1126/science.abq6100} offers a new tool by which to characterize the strange metal, supplementing existing measurements of resistivity and other thermodynamic quantities. It is found in YRS that in  the linear-in-$T$ regime, the shot noise is anomalously small. 


In mesoscopic conductors, fluctuation of electric current is a ubiquitous phenomenon, and the correlation of such fluctuations in time defines the current noise\cite{BLANTER20001,deJong1997,PhysRevB.47.16427,doi:10.7566/JPSJ.90.102001,LANDAUER1991167}. 
There are  two main sources of current noise: thermal noise and shot noise.  The former, also known as Johnson-Nyquist (JN) noise, is essentially an equilibrium property, as it follows from the fluctuation-dissipation theorem (FDT).  This can be seen from the fact that the JN noise is directly proportional to the conductance of the system.  \jjy{The JN noise thus provides information redundant with the average current.}  By contrast, shot noise is inherently a non-equilibrium property \jjy{that exists even in the zero-temperature limit}. It arises from the granularity of charge carriers and varies between systems, offering a unique approach to investigating electronic properties of a conducting system including fractional charge, Dirac fermions or Majorana fermions etc.\cite{de-Picciotto1997,PhysRevLett.79.2526,PhysRevLett.101.120403,PhysRevB.88.064509,PhysRevLett.100.196802}.

\jjy{A simple quantitative description of shot noise is given by a metric called the Fano factor. Consider an ideal two-terminal mesoscopic device at $T=0$, with classical charge carriers transported through the tube in a Poissonian fashion. The resultant shot noise power is given by the Schottky formula $\mathcal{S}_\text{ideal}=2e\braket{I}$ where $e$ is the charge of the carrier and $\braket{I}$ is the average current. However, realistic systems often exhibit shot noise power smaller than this ideal limit\cite{PhysRevLett.75.1610,PhysRevLett.76.3806}\footnote{exceptions include cases where spin-flip second order tunneling dominates\cite{PhysRevLett.95.146806}}. Indeed, just by promoting the charge carriers from classical balls to fermionic electrons, their correlations (due to Fermi-Dirac statistics) will  suppress the shot noise relative to the Poissonian value. The Fano factor $\mathcal{F}$, defined as $\mathcal{F}=\mathcal{S}_\text{shot}/(2e\braket{I})$, quantifies this suppression.}

\jjy{While the Fano factor depends on the specific scattering mechanisms present in a system, there are a few situations in which it is theoretically robust. In particular, the diffusive Fermi liquid with impurity-dominated scattering exhibits $\mathcal{F}=1/3$ \cite{NAGAEV1992103}, independent of the strength of this scattering. On the other hand, a diffusive Fermi liquid with scattering dominated by electron-electron interactions is 
predicted to have a Fano factor of $\mathcal{F}=\sqrt{3}/4$ in the zero temperature limit\cite{PhysRevB.52.4740,PhysRevB.52.7853}. Lastly, in systems with dominant electron-phonon scattering, }the Fano factor vanishes in the limit when the system size becomes large enough, due to efficient equilibration from the phonon bath.

Intriguingly, shot noise in YRS has revealed a Fano factor in the temperature range $1K \lesssim T \lesssim 10K$ (where linear-in-$T$ resistivity is observed) to be significantly smaller than both the characteristic values one obtains by consdering either impurity scattering or electron-electron scattering.  
Furthermore, by studying different sample lengths, the phonon-driven suppression has been ruled out\cite{doi:10.1126/science.abq6100}.   The anomalous suppression of shot noise in this system, and its relation to the strange metallicity, invites us to investigate the shot noise arising from such strange metals. Indeed, there have been several recent theoretical attempts at addressing this suppressed noise in YRS\cite{PhysRevResearch.5.043143, wu2023suppression,wang2024shotnoisecoupledelectronboson,raghu2024shotnoisenearquantumcriticality}.

In this paper, we aim to provide a simple explanation for the observed small Fano factor in a strange metal. Our point of departure, {similar to the spirit of Ref.\cite{raghu2024shotnoisenearquantumcriticality}}, is a phenomenological model of a marginal Fermi liquid\cite{PhysRevLett.63.1996,PhysRevB.44.293,PhysRevB.46.405,10.1063/1.348195}(MFL), where  electron-electron interactions  induce a strongly energy- and temperature-dependent inelastic scattering. 
{The peculiar and appealing feature of MFL is that, although the fermion scattering is inelastic, certain cancellations in the collision integral makes it behave like elastic scattering, and the kinetic equations are reminiscent of those in the impurity scattering case.}
Among many proposals on the microscopic origins of MFL\cite{PhysRevLett.94.156401,PhysRevB.59.R2474,PhysRevLett.124.186801,PhysRevB.90.121107},  
one appealing route is
to couple of the fermions  to a critical  boson, which in turn is a collective mode of the fermions on the verge of condensation\cite{PhysRevB.109.075107,Zhang2023,PAN2024116451,bashan2023tunable,tulipman2024solvable,refId0,PhysRevB.91.045110,PhysRevX.8.031024,RevModPhys.94.035004,PhysRevB.103.235129,PhysRevB.106.115151}. At the level of simple phenomenology, we simply proclaim that the  bosons have a fluctuation spectrum that gives rise to a linear in $T$ scattering rate for the fermions.

To be specific, we consider a system in which the conventional impurity scattering competes with the MFL scattering. At higher $T$, MFL scattering dominates, and the system shows linear-in-$T$ resistivity, while at low $T$ impurity scattering dominates. 
We then investigate the shot noise power in this system based on non-equilibrium quantum field theory which, approximated at the same accuracy level as in the analysis of Refs.\cite{NAGAEV1992103,PhysRevB.52.4740,PhysRevB.52.7853}, gives the same results as the conventional Boltzmann-Langevin approach\cite{Gantsevich1979,kogan1969theory}.

Within a set of assumptions to be described in detail in the next section, we are able to show quite simply the intimate relation between linear-T resistivity of the strange metal and a vanishingly small Fano factor.  Our main results are summarized as follows.   
The total noise power, $\mathcal{S}(T,V)$, in a system with dominant MFL scattering is found to be a $T$-independent and $V$-independent constant\footnote{In this paper we will use $T$ and $T_0$ interchangeably. The latter is preferred when we introduce the local temperature $T(x)$ where $T_0$ is defined the boundary values of $T(x)$, and thus the environmental temperature $T$.}. This constant is exactly given by the noise power when $V=0$, i.e. the JN noise.  
Indeed, since the JN noise is given by $\mathcal{S}_\text{JN}=4GT$, within the assumption of  linear-in-$T$ resistivity regime the conductance $G\propto T^{-1}$ and therefore $\mathcal{S}$ becomes a constant.  As a result, the measured shot noise, defined as $\mathcal{S}_\text{shot}=\mathcal{S}(T,V)-\mathcal{S}(T,0)$ vanishes identically.  From our vantage point, this is a feature that is unique to MFLs; more general non-Fermi liquids would be expected to have a temperature-dependent JN and a non-vanishing shot noise power.  
It is worth commenting that the vanishing of shot noise in a pure MFL is different from that in a system with strong electron-phonon coupling {at $T$ below the its Debye temperature $T_{D}$}: unlike in the case of phonons, in the clean MFL,  the Fano factor does not vanish due to an increase of the system size. 
We also find that, if additional impurity scattering is introduced as a weak perturbation compared to the MFL scattering, the resulting Fano factor reaches a finite value, which presumably accounts for the  experimentally observed Fano factor in YRS.

This paper is organized as follows.
\jjy{In Sec.\ref{sec:shot_noise_in_a_pure_mfl}, we present a general formalism used to calculate transport quantities, including the shot noise. We establish the phenomenological model of the MFL in Sec. \ref{sec:model}, and present the numerical results in Sec. \ref{sec:solutions_of_kinetic_equation}. Some concluding remarks are left in Sec.\ref{sec:discussion_and_conclusion}. All the detailed derivations relevant to the main text are presented and discussed the the Appendices.}

\section{General formalism of kinetic equation and shot noise} 
\label{sec:shot_noise_in_a_pure_mfl}

{Before discussing the MFL model and its shot noise, we describe here the physical setup, the formalism, and the assumptions and approximations that are necessary to obtain our main results.  }

\begin{figure}
    \includegraphics[width=8.cm]{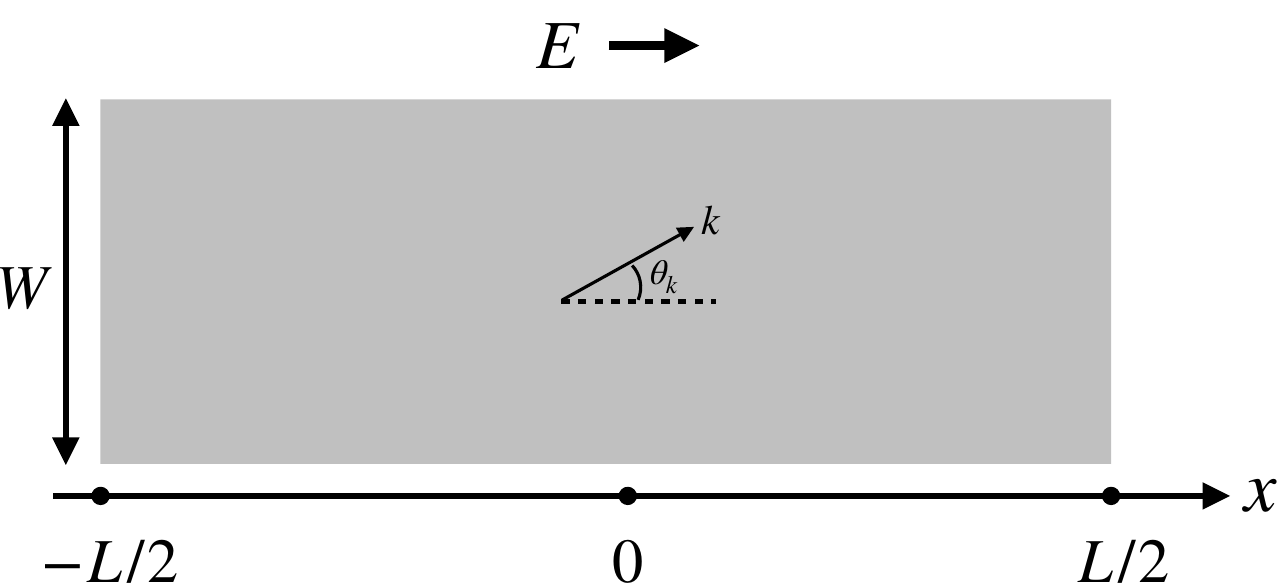}
    \caption{Conducting sample under consideration, which has a length $L$ and width $W$. The applied electric field is along the $x$-direction. }\label{fig:sample}
\end{figure}

\subsection{Setup and definitions}
\jjy{We consider a typical transport experiment: a rectangular sample is attached to {two leads at $x=\pm L/2$}, 
and current is induced by a voltage difference between the leads. Concretely, we model the sample as} a slab geometry which can be effectively viewed as a two-dimensional system with length $L$ and width $W$, as shown in Fig. \ref{fig:sample}. The electric field is applied along the $x$-direction, so $\bm{E}=\hat x E$. 

\jjy{The central quantity in our approach is the distribution function $F$ of the charge carriers\footnote{In our convention, the distribution function $F$, in the equilibrium limit, reduces to $1-2n_F$ where $n_F$ is the usual Fermi-Dirac distribution}, which will ultimately be used to calculate physical observables.} 
\jjy{Away from equilibrium, $F$ must be determined by solving the kinetic equation. In the fully quantum-mechanical treatment of kinetic theory, the distribution function formally depends on position $\mathbf{r}$, momentum $\mathbf{k}$, frequency $\omega$, and time $t$. The {kinetic (quantum Boltzmann)} equation has the general form 
(see Appendix \ref{sec:appB} for details): 
\begin{equation}
  \begin{aligned}
        &\left(Z^{-1}\partial_t+\bm{v}\cdot\nabla_{\br}-\nabla_{\br}\tilde{V}\cdot\nabla_{\bk}+\partial_t\text{Re}\Sigma^R\partial_\omega\right)F(\br,t,\bk,\omega)\\
        &=I_\text{col}(\br,t,\bk,\omega)
   \end{aligned} 
    \label{eq:boltzmann_general}
\end{equation}
where
\begin{equation}
    \begin{aligned}
        Z^{-1}&=1-\partial_\omega\text{Re}\Sigma^R(\br,t,\bk,\omega),\\
        \bm{v}_{\bk}&=\nabla_{\bk}\epsilon_{\bk}, ~ \epsilon_{\bk}=\epsilon_{\bk}^{(0)}+\text{Re}\Sigma^R(\br,t,\bk,\omega),\\
        \tilde{V}&=V(\br)+\text{Re}\Sigma^R(\br,t,\bk,\omega),
    \end{aligned}
\end{equation}
are the inverse quasiparticle residue, renormalized velocity and renormalized potential respectively. They are all determined by the retarded fermion self energy $\Sigma^R(\br,t,\bk,\omega)$, which, in turn, also depends on $F$ in a self-consistent way (as will be shown below, in our MFL model we  neglect the momentum dependence of the fermion self energy).
$V(\br)$ is the bare external potential that incorporates the electric field and  $\epsilon_{\bk}^{(0)}$ is the bare quasiparticle dispersion relative to the chemical potential. The collision integral $I_{\text{col}}$ incorporates all scattering mechanisms in the theory, which will be explicitly shown below.
}

\jjy{Solving the Boltzmann equation in its full form is a hopeless endeavor, even with moderate Fermi liquid-like scattering. For non-Fermi liquids, additional complexity arises due to the singular fermion self energy. For instance, the quasiparticle residue $Z$ vanishes at $\omega\to0$ and hence $Z^{-1}$ in the Boltzmann equation becomes singular at $\omega=0$, which requires additional care. In the next section, we will describe the various assumptions that allow us to find a solution within a semiclassical approximation and the caveats of using such a solution. }

\subsection{Caveats and assumptions }

\jjy{While a rigorous solution of the Boltzmann equation is impossible, we can make a series of assumptions to simplify the task.  Here, we reiterate these common assumptions in the interest of clarity. Having disposed of these caveats, we then proceed with the results. }

First of all, we will be interested in steady state solutions and will neglect the time-dependence of the distribution function. In this case, the singular behavior of $Z^{-1}$ can be excluded. \jjy{This is a reasonable approximation akin to neglecting any transient behavior of the system in short times after the voltage bias has been applied.}

Our next simplifying approximation is to neglect self-energy effects on the single-particle lifetime. This is usually dangerous for a more violent non-Fermi liquid, but for the MFL (as we will justify below), restricting our attention { to  Fermi surface with such an approximation can indeed produce} linear-in-$T$ resistivity, and we thus study the current noise within the same framework. 
Mathematically, the approximation is reflected in the ``on shell" condition for the distribution function, namely
\begin{equation}
F(\br,\bk,\omega)\propto F(\br,\bk)\delta(\omega-{\epsilon^{(0)}_{\bk}}).\label{eq:onshell}
\end{equation}
\jjy{It is precisely this assumption that allows us to proceed with a semiclassical approach to solving the Boltzmann equation and eventually leads to a simple interpretation of the collision integral.}

\jjy{Under these two assumptions, the} Boltzmann steady-state kinetic equation governing the simplified distribution function $F(\br,\bk)$ is
\begin{equation}
(\bm{v}_{\bk}\cdot\nabla_{\br}+ \mathbf{E}\cdot \nabla_{\bk})F(\br,\bk)=I_\text{col}(\br,\bk)\label{eq:BE1},
\end{equation}
where $\bm{v}_{\bk}$ is the velocity. We have further assumed that the position dependence for the self energy is weak here, such that only the electric field contributes to the potential.

Depending on the size of the system and the strength of the electric field, there exist two practical ways to proceed. If the system is large and the field is weak, as in the regime where linear response theory applies, one argues that the spatial dependence of the distribution function is weak, and thus the first term $\bm{v}_{\bk}\cdot\nabla_{\br}$ can be dropped\cite{lifshitz1995physical,abrikosov2017fundamentals,mahan2013many,bruus2004many}. However, in mesoscopic systems with large applied field, the fermion chemical potential can be better understood as a  local quantity $\mu(\br)$ whose gradient is the electric field. The $\bm{k}$-dependence of $F(\br,\bk)$ through $\epsilon_{\bk}$ can now be understood as dependence on $\epsilon_{\bk}-\mu(\br)$. Therefore, the $\bm{E}\cdot\nabla_{\bk}$ operator can be absorbed into the first term $\bm{v}_{\bk}\cdot\nabla_{\br}$ if the spatial gradient is interpreted as a total derivative. In this way $\bm{E}\cdot\nabla_{\bk}$ is effectively dropped\cite{Frederick_Green_2000,PhysRevB.59.13054,PhysRevB.60.5839,kamenev2023field}. We adopt the second approach, as it is more relevant to the shot noise problem.  We also note from our calculations that in both approaches, keeping both $\bm{v}_{\bk}\cdot\nabla_{\br}$ and $\bm{E}\cdot\nabla_{\bk}$ simultaneously leads to incorrect transport coefficients due to double counting. 

Finally, we make a few assumptions related to the setup of the system. First, we shall ignore dependence on spatial components perpendicular to $\bm{E}$. For our purposes (see  Fig. \ref{fig:sample}), this means that we assume the distribution function is invariant along the $y$ direction. Second, we assume that the small voltage bias is small enough (compared to the Fermi energy) such that the only charge carriers contributing to transport are close to the Fermi energy, so $v_k = v_F$. 
With the assumptions described above, the Boltzmann equation takes the simpler form
\begin{equation}
v_F \frac{\partial F(\br, \bk)}{ \partial x} = I_\text{col}(\br,\bk)\label{eq:BE1}.
\end{equation} 
subject to some proper boundary conditions which will be specified below.

\subsection{Strategy and notation}
Before embarking on applying the above mentioned formalism to our MFL model, 
here we detail our strategy for solving the (simplified) Boltzmann equation \eqref{eq:BE1} and for computing the relevant physical observables. We first introduce
the partial wave decomposition\cite{mahan2013many} to the distribution function, namely, 
\begin{equation}
    F(\br,\bk)=F_s(x,\epsilon)+\cos\theta_{\bk}F_p(x,\epsilon)+...\label{eq:partialwave}
\end{equation}
where $\epsilon=\epsilon_{\bk}$ and all higher order partial waves are omitted. In this rewriting, we invoke the assumption that the system is invariant along the $y$-direction (perpendicular to the applied field), so $F_s$ and $F_p$ depend only on $x$. 
The $s$-wave component $F_s$ preserves inversion ($\bk \rightarrow -\bk$), while the $p$-wave part $\cos\theta_{\bk}F_p$ does not. Since the current in the $x$-direction by definition breaks this inversion symmetry,  it can only depend on $F_p$.

With this decomposition, the Boltzmann equation can be rewritten as two coupled equations. To this end,  we substitute Eq.\eqref{eq:partialwave} into Eq.\eqref{eq:BE1}, and
i) integrate over all $\theta_{\bk}$ and ii) multiply Eq.\eqref{eq:BE1} by $\cos\theta_{\bk}$ and then integrate over $\theta_{\bk}$. The resulting equations are 
\jjy{
\begin{equation}
    \begin{aligned}
        v_F\partial_xF_p(x,\epsilon)&=I_{\text{col},s},\\
        v_F\partial_xF_s(x,\epsilon)&=I_{\text{col},p},
    \end{aligned}\label{eq:BE2}
\end{equation}
where the collision integrals are decomposed into partial wave components as well. This decomposition also makes clear the effects of different parts of the collision integral. In particular, $I_{\text{col},s}$ corresponds to energy relaxation while $I_{\text{col},p}$ corresponds to momentum relaxation. In Appendix \ref{sec:appB}, we provide a general formalism from which collision integrals due to various scattering mechanisms can be derived. } 

We will use the so-called relaxation time approximation to express $I_{\text{col},p}$ in terms of a characteristic relaxation rate $\tau(x,\epsilon)$
\begin{equation}
   I_{\text{col},p} = -\frac{1}{\tau(x,\epsilon)} F_p(x,\epsilon),\label{eq:relaxtimeapprox}
\end{equation}
In fact, as will be shown,  this approximation can be derived when some proper ansatz for $F_s$ has been used. 
The scattering time $\tau(x,\epsilon)$ is generically defined in terms of the distribution function and  ultimately encodes the important information about various scattering mechanisms in the system. Under the assumption that there is no interference of scattering processes, the collision integrals and therefore scattering rates of different mechanisms should add.

There are two main ansatzes for the distribution function which are commonly used to solve the Boltzmann equation: the ``local-temperature ansatz" and the ``diffusion ansatz." The first is a distribution function which takes the equilibrium form but with a spatially-varying temperature profile\cite{PhysRevB.52.4740}: 
\begin{equation}
F_s(x,\epsilon)=\tanh\frac{\epsilon+|e|E x}{2T(x)}\label{eq:ansatz1}.
\end{equation} 
{This approach is justified when the dominant scattering is inelastic, and the scattering length is much smaller compared to the system size $L$.}
The second ansatz applies 
{when the scattering is dominated by elastic scattering, like in the strong impurity scattering case,}
where the dynamics are diffusion-dominated\cite{NAGAEV1992103}: 
\begin{equation}
F_s(x,\epsilon)=\frac{1}{2}\left[F_R(\epsilon)+F_L(\epsilon)\right]+\frac{x}{L}\left[F_R(\epsilon)-F_L(\epsilon)\right]\label{eq:ansatz2}.
\end{equation} 
{In both ansatz, we will impose the following boundary condition 
\begin{equation}
    F_s(\pm L/2,\epsilon)=\tanh\frac{\epsilon\pm |e|V/2}{2T(x)}
\end{equation}}
In the first ansatz, the specific scattering mechanisms (encoded in the collision integral) will determine the temperature profile $T(x)$; in the second ansatz, they will determine the mean free path.  Below we will use Eq.\eqref{eq:ansatz1} for case where MFL scattering dominates, and use Eq.\eqref{eq:ansatz2} for the case where impurity scattering dominates.

\jjy{Regardless of which ansatz we use to find the distribution functions $F_s(x,\epsilon)$ and $F_p(x,\epsilon)$, we can directly compute the transport properties once these functions are known.}
Our formalism of computing the average current and noise power is based on the non-equilibrium field theory, which is different from the Boltzmann-Langevin approach but yields the same results (see Appendix \ref{sub:fluctuations_above_the_saddle_point} for more details). 
For instance, the local current density is given by
\begin{equation}
    \braket{\bm{j}(x)}=-\frac{|e|}{2}\sum_{\bk}F(\br,\bk)\bm{v}_{\bk}.
\end{equation}
Here $\nu_0$ is the density of states at Fermi level. 
Note that although $F(\br,\bk)$ contains both $s$-wave and $p$-wave components, as shown in Eq.\eqref{eq:partialwave}, only $F_p(x,\epsilon)$ leads to nonzero current. 
{The total current is given by $\braket{I}=\frac{W}{L}\int_{-L/2}^{L/2}dx\braket{j_x(x)}$.}
It is then straightforward to obtain the total average current 
\begin{equation}
    2e\braket{I}=-\frac{e^2v_F^2\nu_0W}{2L}\int_{-\frac{1}{2}}^{\frac{1}{2}}\frac{dx}{L}\int_{-\infty}^{+\infty}\frac{d\epsilon L}{v_F}F_p(x,\epsilon).\label{eq:aveI}
\end{equation}
The total noise power for a fixed boundary temperature $T_0\equiv T(x=\pm L/2)$ and voltage V, is defined and calculated as (see Appendix \ref{sub:current_noise} for details)  
\begin{equation}
    \begin{aligned}
        &\mathcal{S}(T_0,V)=2\int_0^{+\infty} dt \braket{\delta I(t)\delta I(0)}\\
        &\approx\frac{e^2v_F^2\nu_0W}{2L}\int_{-\frac{1}{2}}^{\frac{1}{2}}\frac{dx}{L}\int_{-\infty}^{+\infty}d\epsilon[1-F_s^2(x,\epsilon)]\tau(x,\epsilon).
    \end{aligned}\label{eq:noisepower1}
\end{equation}
Here $\delta I(t)= I(t)-\braket{I}$, so the noise power is nothing but the connected correlations of the total current operator, averaged over time\footnote{In the language of field theory, it is identical to the second order functional derivative of $\ln Z[\bm{A}]$ (as opposed to $Z[\bm{A}]$) with respective to the external gauge potential $\bm{A}$, where $Z[\bm{A}]$ is the generating function. }.  To arrive at the second line Eq.\eqref{eq:noisepower1}, we only keep the leading contribution, which is at the same accuracy level as that in the bulk literature on shot noise. In particular, if $\tau$ is  a constant, this expression reduces to the the one derived from Boltzmann-Langevin (BL) formalism in Ref.\cite{NAGAEV1992103,PhysRevB.52.7853,Gantsevich1979}.

This total noise power contains both JN noise and shot noise. As we discussed in the Introduction, the JN noise is an equilibrium property, which has to be extracted from the total noise. The experimentally measured shot noise is therefore given by 
\begin{equation}
    \mathcal{S}_\text{shot}(T_0,V)=\mathcal{S}(T_0,V)-\mathcal{S}(T_0,0),
\end{equation}
from which the Fano factor is readily obtained.


\section{Phenomenological model of a marginal Fermi liquid}
\label{sec:model}

\jjy{Having established the formalism, we now focus on applying it to the MFL model to explain the observations reported in \cite{doi:10.1126/science.abq6100}.} 
\jjy{Below, we describe this model and report its salient features, including the intermediate results necessary for computing the Fano factor.}

\subsection{Model of MFL}

In the phenomenological MFL model, fermions  couple to some bosonic mode whose spectral function exhibits frequency/temperature scaling:
\begin{equation}
    B(\bq,\Omega)=\frac{1}{\pi\omega_D}\tanh\frac{\Omega}{2T}\label{eq:MFLspectral}
\end{equation}
where $\omega_D$ is a characteristic frequency associated with the bosonic fluctuation spectrum. 
We assume that the notion of ``local quantum criticality" and neglect any singular momentum dependence of the boson fluctuation spectrum.  As fermions near the Fermi surface scatter off of these bosonic modes, they acquire some nontrivial self energy correction even at the one-loop level. If we identify the imaginary part of the retarded self energy $\text{Im}\Sigma^R(\epsilon)$ as $\frac{1}{2\tau_{\text{MFL}}}$, it is easy to see (as shown in Appendix \ref{sec:MFLselfenergy})
\begin{equation}
    \tau_{\text{MFL}}(\epsilon)=\frac{1}{2g^2}\frac{1}{\epsilon}\tanh\frac{\epsilon}{2T}  \label{eq:MFLparticlelifetime}
\end{equation}
where $g$ is the effective MFL coupling strength. 

At low frequencies $\epsilon \ll T$, the quasiparticle scattering rate $ \tau_{\text{MFL}}^{-1} \sim T$ is comparable to the quasiparticle energy, and long-lived quasiparticles cease to exist. \jjy{Though $\tau_{\text{MFL}}$ is a single-particle property, we will soon find that the transport scattering rate $\tau_{\text{M}}$ due to the MFL interaction  is closely related.}

In addition to the MFL scattering, we also assume some impurity scattering from a local disorder $V_\text{dis}(\br)$, whose correlation is $\overline{V_\text{dis}(\br)V_\text{dis}(\br')}=g(\br-\br')$. In the self-consistent Born approximation, this disorder potential, after disorder average, gives rise to some finite self energy $i\frac{1}{2\tau_\text{im}}$, where the impurity scattering time $\tau_\text{im}$ is determined by the correlation $g(\br)$. The same $\tau_\text{im}$ also determines the transport scattering rate that enters the collision integral. 

\subsection{Collision integrals and scattering time in the MFL}

In the presence of both MFL scattering and impurity scattering, the total collision integral captures both effects, namely we have $I_\text{col}=I_\text{im}+I_\text{MFL}$. We leave the explicit forms of $I_\text{im}$ and $I_\text{MFL}$ in Appendix \ref{sec:appB}, which, after the partial wave decomposition, become $I_{\text{col},s}=I_{\text{im},s}+I_{\text{MFL},s}$ and $I_{\text{col},p}=I_{\text{im},p}+I_{\text{MFL},p}$ where
\begin{equation}
    I_{\text{im},s}=0,~~ I_{\text{im},p}=-\frac{1}{\tau_\text{im}} F_p(x,\epsilon),\label{eq:Iim}
\end{equation}
and 
\begin{equation}
    \begin{aligned}
        I_{\text{MFL},s}&=g^2\int d\epsilon' \left\{[F_s(x,\epsilon')-F_s(x,\epsilon)]\frac{}{}\right.\\
        &~~~~~~~~~~~~\left.-[1-F_s(x,\epsilon)F_s(x,\epsilon')]\tanh\frac{\epsilon'-\epsilon}{2T(x)}\right\},\\
         I_{\text{MFL},p}&=g^2\int d\epsilon' \left[F_s(x,\epsilon')\tanh\frac{\epsilon'-\epsilon}{2T(x)}-1\right]F_p(x,\epsilon).
    \end{aligned}\label{eq:IMFLsp}
\end{equation}
In deriving $I_\text{MFL}$ we have assumed that the bosons are also in local equilibrium such that the same local temperature $T(x)$ defined for the fermions and also be defined for the bosons. 

If the MFL scattering is turned off, i.e. $g=0$, the system becomes a conventional dirty metal. In this case, inserting Eq.\eqref{eq:Iim} into Eq.\eqref{eq:BE2} yields a simple diffusion equation for the $s$-wave component, namely $\partial_x^2F_s(x,\epsilon)=0$. The solution subject to proper boundary condition is exactly the second ansatz that we have discussed in Eq.\eqref{eq:ansatz2}. On the other hand, if we keep only the MFL scattering by setting $\tau_\text{im}=\infty$, we will solve Eq.\eqref{eq:BE2}{} with the first ansatz in Eq.\eqref{eq:ansatz1}. The advantage of the local temperature ansatz in this case becomes manifest after we substitute it into Eq.\eqref{eq:IMFLsp} and perform the integral,
\begin{equation}
    I_{\text{MFL},s}=0,~~ I_{\text{MFL},p}=-\frac{1}{\tau_\text{M}(x,\epsilon)} F_p(x,\epsilon),\label{eq:IMFL}
\end{equation}
where the MFL transport time is 
\begin{equation}
    \tau_\text{M}(x,\epsilon)=\frac{1}{2g^2(\epsilon + |e|Ex)}\tanh\frac{\epsilon + |e|Ex}{2T(x)}.\label{eq:tauMFL}
\end{equation}
{Eq.\eqref{eq:IMFL} is remarkable in the sense that, even if we started with some elastic scattering in which fermions indeed transfer energy, the resulting $I_{\text{MFL},s}$ still vanishes. In this way, we may think of the MFL scattering as sort of a temperature- and energy-dependent  elastic impurity scattering.} Moreover, comparing Eq.\eqref{eq:tauMFL} to Eq.\eqref{eq:MFLparticlelifetime} we see that $\tau_\text{M}(x,\epsilon)$ can be simply obtained by replacing $\epsilon$ in $\tau_\text{MFL}(\epsilon)$ with $\epsilon+|e|Ex$ and promoting $T$ to $T(x)$. 
The formal similarity between Eq.\eqref{eq:IMFL} and \eqref{eq:Iim} allows for the definition of the total transport time
\begin{equation}
    \tau^{-1}(x,\epsilon)=\tau_\text{M}^{-1}(x,\epsilon)+\tau_\text{im}^{-1},\label{eq:totalTau}
\end{equation} 
such that $I_{\text{col},p}=-F_p(x,\epsilon)/\tau(x,\epsilon)$, which coincides with the relaxation time approximation commonly used for deriving Drude conductivity in a conventional dirty metal, and thus justifies our previous assumption in Eq.\eqref{eq:relaxtimeapprox}

In fact, $I_{\text{col},s}=0$ as inferred from Eq.\eqref{eq:Iim} and \eqref{eq:IMFL} is a very strong statement, which holds only if we accept the local temperature ansatz for $F_s(x,\epsilon)$ and assume the bosons are also in local equilibrium with the same $T(x)$. Moreover, we have not included any residual fermion-fermion interaction that can lead to inelastic scattering. Violating any of these conditions would lead to a finite $I_{\text{col},s}$. Therefore, the more cautious treatment for $I_{\text{col},s}$, especially in the regime where MFL scattering dominates, is to assume local hydrodynamics with the conservation of particle number and energy. Specifically we will use the following two conditions for the MFL scattering dominated regime\cite{PhysRevB.52.4740},
\begin{equation}
    \int_{-\infty}^{+\infty} d\epsilon I_{\text{col},s}(x,\epsilon)=0,~ \int_{-\infty}^{+\infty} d\epsilon \epsilon I_{\text{col},s}(x,\epsilon)=0,\label{eq:hydro}
\end{equation}
for every spatial position $x$.

\section{Results} 
\label{sec:solutions_of_kinetic_equation}

We now present the solutions of the Boltzmann equation \eqref{eq:BE2} and the resulting shot noise based on the formalism outlined above. Depending on the relative strength between the MFL scattering and the impurity scattering, three different cases will be discussed in order: i) the clean limit with only MFL scattering dominates, ii) when both MFL scattering and impurity scattering are present but the latter is treated as perturbation and iii) impurity scattering dominates and MFL scattering is treated as perturbation. 


\subsection{Clean MFL} 
\label{sub:clean_mfl}


\begin{figure}
    \includegraphics[width=8cm]{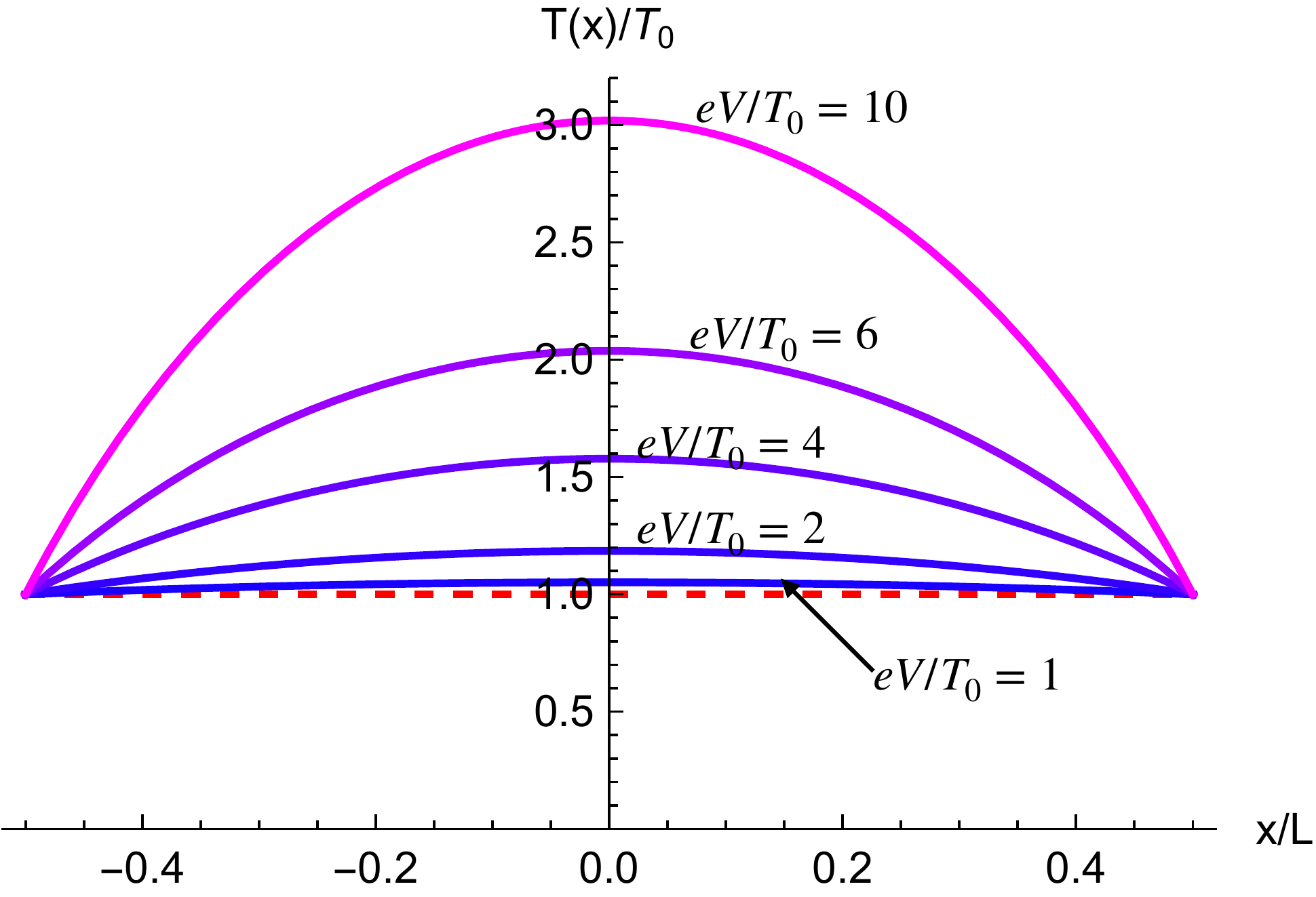}
    \caption{Solution of the differential equation in \eqref{eq:diffTx} for various ratios of $eV/T_0$.}\label{fig:Tx1}
\end{figure}

We first discuss the case when $\tau_\text{im}=\infty$ such that $\tau(x,\epsilon) = \tau_{\text{M}}(x,\epsilon)$. Using Eq.\eqref{eq:IMFLsp} to eliminate $F_p(x,\epsilon)$ in Eq.\eqref{eq:BE2}, and by applying both of the hydrodynamic conservation laws for $I_{\text{col},s}$ in Eq.\eqref{eq:hydro}, we obtain
\begin{equation}
    \int_{-\infty}^{+\infty}dy y \partial_x[\tau_\text{M}(x,\epsilon)\partial_xF_s(x,\epsilon)]=0,
\end{equation}
where we have introduced $y=\epsilon+|e|Ex$ for convenience. Inserting $F_s(x,\epsilon)$ and $\tau_\text{M}(x,\epsilon)$ from Eq.\eqref{eq:ansatz1} and \eqref{eq:tauMFL} and performing the integral, we obtain the differential equation for $T(x)$,
\begin{equation}
   L^2 \frac{\partial_x^2T(x)}{T_0}+\beta\frac{T_0}{T(x)}=0,\label{eq:diffTx}
\end{equation}
where $\beta=\frac{7\zeta(3)}{2\pi^2}\left({eV}/T_0\right)^2$ and $\zeta(...)$ is the Riemann zeta function. 
The formal solution consistent with the boundary condition can be expressed in terms of the inverse error function InvErf, 
\begin{equation}
    T(x)=T_0\exp\left[\xi-\text{InvErf}^2\left(\sqrt{\frac{2\beta x^2}{\pi L^2}e^{-2\xi}}\right)\right]\label{eq:formalTx}
\end{equation}
with $\xi$ from the solution of the boundary condition equation
\begin{equation}
    \text{Erf}(\sqrt{\xi})=\sqrt{\frac{\beta}{2\pi}e^{-2\xi}}.
\end{equation}

In Fig.\ref{fig:Tx1} we plot the numerical solution of $T(x)$ for various ratios of $eV/T_0$. We see that $T(x)$ is a symmetric function of $x$, and as the bias increases from zero, $T(x=0)$ deviates from $T_0$ more and more drastically. In fact, it is not hard to see that as $T_0\to0$, $T(x)$ is set only by the voltage bias $V$. However, the ansatz in Eq.\eqref{eq:ansatz1} is is invalid in this limit. Recall that we must have $\tau_\text{M} \ll \tau_\text{im}$ to justify the use of Eq.\eqref{eq:ansatz1}, but as $T_0\to0$, $\tau_\text{M}(x,\epsilon)$ at energies $\epsilon\sim|e|Ex$ becomes infinitely large, especially near the boundaries.
Therefore, we argue that the above results are reliable only when $T_0$ is not so small. 

The peculiar property of MFL can be seen from the particular energy dependence of $\tau_\text{M}$, i.e. the inverse proportionality to $y=\epsilon+|e|Ex$. From Eq.\eqref{eq:noisepower1} we find that the total noise power is given by
\begin{equation}
    \begin{aligned}
        \mathcal{S}(T_0,V)&=\frac{e^2v_F^2\nu_0W}{4g^2L}\int_{\frac{-L}{2}}^{\frac{L}{2}}\frac{dx}{L}\\
        &\times\int_{-\infty}^{+\infty}\frac{dy}{y}\left[1-\tanh^2\frac{y}{2T(x)}\right]\tanh\frac{y}{2T(x)}\\
        &=\frac{7\zeta(3)}{\pi^2}\frac{e^2v_F^2\nu_0W}{2g^2L}=\frac{W}{L}\frac{7\zeta(3)}{g^2\pi^2}\frac{ne^2}{m}.
    \end{aligned}\label{eq:noisepower2}
\end{equation}
where $\nu_0$ is the density of states near Fermi level, $m$ is the effective mass and $n$ is the fermion density. 
The fact that the total noise power depends on neither $T_0$ nor $V$ directly indicates that 
\begin{equation}
    \mathcal{S}_\text{shot}=\mathcal{S}(T_0,V)-\mathcal{S}(T_0,0)=0.\label{eq:MFLshot}
\end{equation}
The vanishing of shot noise is purely due to the specific energy dependence of $\tau_\text{M}$ and is therefore is a unique property of a MFL. Interestingly, the specific form of $T(x)$ does not matter here: a different $T(x)$ gives the same result in Eq.\eqref{eq:MFLshot} as long as the system is not in the zero-temperature limit\cite{raghu2024shotnoisenearquantumcriticality}. Therefore, we expect that in the clean strange metal regime where MFL scattering dominates, the shot noise (and hence the Fano factor) should vanish. Another implication of Eq.\eqref{eq:MFLshot} is that the JN noise is a constant, consistent with the expression $\mathcal{S}_\text{JN}=4GT$ and $G\sim T^{-1}$. This suggests that thermal fluctuations in a strange metal are substantial even at low temperatures. 
Indeed, the average current can be calculated from Eq.\eqref{eq:aveI}, and the result can be neatly written as 
\begin{equation}
    \begin{aligned}
        \braket{I}
        &=\frac{W}{L}\frac{7\zeta(3)}{4\pi^2g^2}\frac{n e^2}{m}  \frac{V}{\overline T} = G V.
    \end{aligned}
    \label{eq:avgcurrent-clean}
\end{equation}
where the average temperature is 
\begin{equation}
    \frac{1}{\overline T}=\int_{\frac{-L}{2}}^{\frac{L}{2}} \frac{dx}{L}\frac{1}{T(x)}.
\end{equation}
If $T(x)\approx T_0$, which is valid when $T_0$ is not very small compared to $eV$, then the average temperature is close to the bath temperature $\overline T\sim T_0$. We see from above that $G\sim 1/T_0$, i.e. the resistivity is linear-in-$T$, and as a result, and the noise power from Eq.\eqref{eq:noisepower2} indeed becomes the JN noise. 

\subsection{MFL with weak impurities} 
\label{sub:mfl_with_weak_impurities}

\begin{figure}
    \includegraphics[width=8cm]{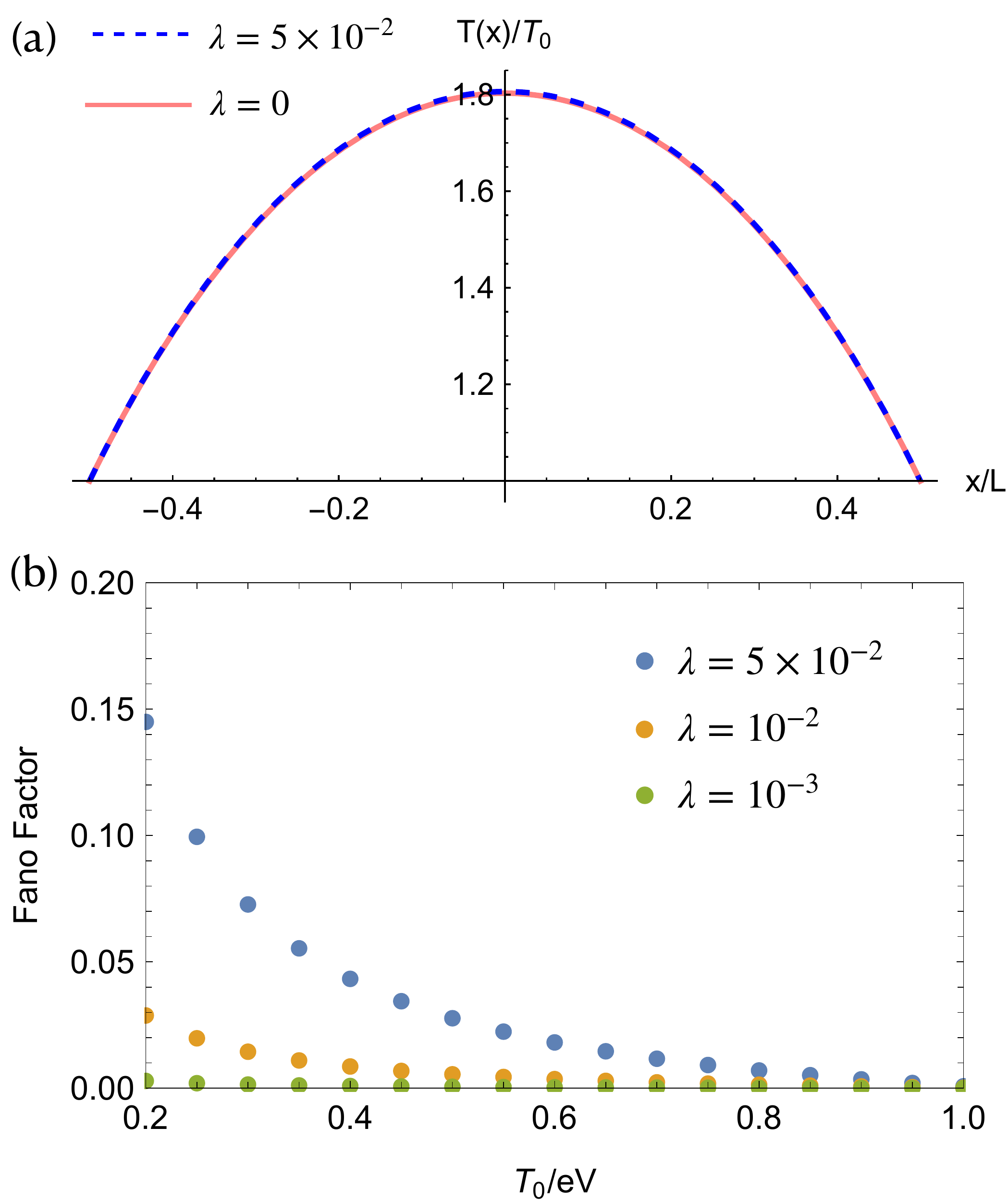}
    \caption{(a) Numerical solution of $T(x)$ in the presence of weak impurities at $T_0/eV=0.2$. Here we remind that $\lambda$ is defined as  $\lambda=1/[g^2\tau_\text{im}eV]$. (b) Fano factors for different impurity strengths.}\label{fig:FanoB}
\end{figure}

Having demonstrated the vanishing of the shot noise in a clean MFL system, we now proceed with the discussion when impurity scattering is included but still subdominant compared to MFL scattering. We thus treat impurity scattering  perturbatively.  
In the regime that $1/[g^2\tau_\text{im}T(x)]$ is a small parameter, we can expand Eq.\eqref{eq:totalTau} as
\begin{equation}
    \tau(x,\epsilon)=\tau_\text{M}(x,\epsilon)-\tau_\text{im}^{-1}\tau_\text{M}^2(x,\epsilon).\label{eq:apprxtau}
\end{equation}
Following the derivations above, we assume the ansatz for the distribution is still given by Eq.\eqref{eq:ansatz1}, but we expect a different solution for $T(x)$. The equation for $T(x)$ is obtained in a parallel way, namely by replacing $\tau_\text{M}$ in Eq.\eqref{eq:diffTx} with the approximate $\tau(x,\epsilon)$ in Eq.\eqref{eq:apprxtau}. After performing the integral, we obtain
\begin{equation}
    \begin{aligned}
        &\partial_x^2 T(x)+\frac{7\zeta(3)}{2\pi^2}\frac{(eE)^2}{T(x)}\\
        &-\frac{1}{6g^2\tau_\text{im}T(x)}\left[\partial_x^2 T(x)-\frac{(\partial_x T(x))^2}{T(x)}+0.56\frac{(eE)^2}{T(x)}\right]=0
    \end{aligned}\label{eq:diffTx2}
\end{equation}
where the numerical factor $0.56$ is from the evaluation of integrals involving hyperbolic functions.  
Note that in the limit when $1/[g^2\tau_\text{im}T(x)]$ vanishes, Eq.\eqref{eq:diffTx2} reduces to Eq.\eqref{eq:diffTx}.

We solve Eq.\eqref{eq:diffTx2} numerically by fixing the external voltage and varying $T_0$. Since $T_0$ is always taken to be some fraction of $eV$, we define the small parameter to be $\lambda=1/[g^2\tau_\text{im}eV]$. 
Fig.\ref{fig:FanoB}(a) shows a comparison between $T(x)$ with and without impurities at $T_0/eV=0.2$. The impurity strength is modeled by $\lambda=0.05$ such that $1/[g^2\tau_\text{im}T(x)]$ can be as large as $0.25$. It is clear from the result that $T(x)$ is barely modified from that in the clean limit, where the formal solution is given by Eq.\eqref{eq:formalTx}.

In the presence of impurities, the average current also changes as a result of the total $\tau(x,\epsilon)$. A straightforward calculation shows that the average current in this case can be obtained by multiplying Eq.\eqref{eq:avgcurrent-clean} by a factor, namely
\begin{equation}
    \braket{I}=\frac{W}{L}\frac{7\zeta(3)}{4\pi^2g^2}\frac{n e^2}{m}\frac{V}{\overline T}\left(1-\frac{0.22}{g^2\tau_\text{im}}\frac{\overline T}{\overline{T^2}}\right).
\end{equation}
In the limit $1/[g^2\tau_\text{im}T(x)]\to0$, this reduces to Eq.\eqref{eq:aveI} as expected.

The Fano factors $\mathcal{F}$ can also be obtained in a straightforward numerical evaluation. The calculation of noise power is still based on Eq.\eqref{eq:noisepower1}, with the total scattering time $\tau(x,\epsilon)$ given in the perturbative expansion in Eq.\eqref{eq:apprxtau}. The results are presented in Fig.\ref{fig:FanoB}(b), for different choices of the dimensionless parameter $\lambda$. In general, $\mathcal{F}$ is a monotonically decaying function of $T_0$. As the impurity strength $\lambda$ decreases, $\mathcal{F}$ quickly decays and eventually becomes zero, which is consistent with Eq.\eqref{eq:MFLshot} in the clean limit above. We argue that the zero temperature limit cannot be obtained in perturbative calculation, since $T_0\to0$ invalidates the expansion in Eq.\eqref{eq:apprxtau} at energies close to $|e|Ex$. In fact, as we discussed above, even in the clean limit, the ansatz \eqref{eq:ansatz1} becomes invalid when $T_0\to0$. In this limit, impurity scattering is the most important, so one must begin with a different ansatz for the distribution function, as we will show below.


\subsection{Strong impurities with weak MFL scattering} 
\label{sub:strong_impurities_with_weak_mfl_scattering}

\begin{figure}
    \includegraphics[width=8cm]{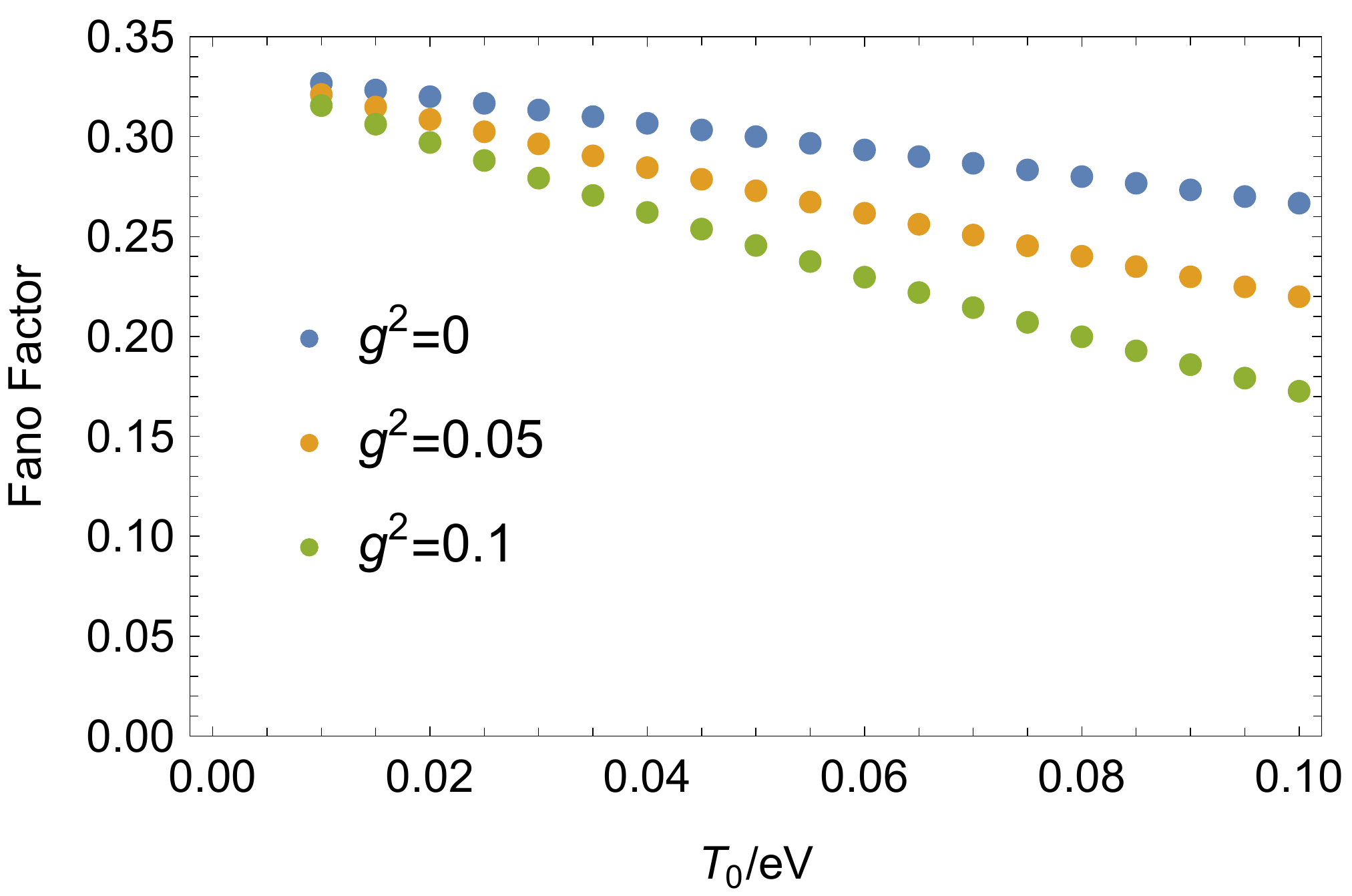}
    \caption{Fano factor in zero temperature limit when the MFL scattering is treated perturbatively. Here $g^2$ is the dimensionless MFL coupling strength, and we take a presentative $\tau_i eV=0.1$, $v_F\tau_i/L=0.1$. }\label{fig:FanoC}
\end{figure}

This regime applies in the zero-temperature limit, when the impurity scattering is the strongest. We also assume the electron-electron scattering is small, so that without MFL scattering, the Fano factor eventually saturates to $1/3$ at $T_0=0$. Therefore, the unperturbed solution for $F_s$, which we denote by $F_s^{(0)}$ is the \jjy{diffusion ansatz Eq.\eqref{eq:ansatz2}.}

When weak MFL scattering is present (the dimensionless $g^2$ is small), we expand the the solution as
\begin{equation}
    F_{s,p}(x,\epsilon)=F_{s,p}^{(0)}(x,\epsilon)+g^2F_{s,p}^{(1)}(x,\epsilon)+\mathcal{O}(g^4).\label{eq:Fperturb}
\end{equation}
To find $F_s^{(1)}$ and $F_p^{(1)}$, we substitute this expansion to Eq.\eqref{eq:BE2} with $I_\text{MFL}$ given in Eq.\eqref{eq:IMFLsp} [but with a constant $T_0$ instead of $T(x)$] and $I_{ee}$ neglected. Matching the coefficients of $g^2$, we obtain 
\begin{equation}
    \begin{aligned}
        F_s^{(1)}(x,\epsilon)&=(4x^2-L^2)\left(\frac{A(\epsilon)}{8}+\frac{B(\epsilon)}{48 L^2}(4x^2-5L^2)\right),\\
        \frac{F_p^{(1)}(x,\epsilon)}{v_F}&=\frac{\tau_\text{im}^2}{L}\left[F_R(\epsilon)-F_L(\epsilon)\right]\frac{2\epsilon\sinh\frac{\epsilon}{T_0}-eV\sinh\frac{eV}{2T_0}}{\cosh\frac{\epsilon}{T_0}-\cosh\frac{eV}{2T_0}}\\
        &~~~~-\frac{(4x^3-3L^2x)\tau_\text{im}}{3L^2}B(\epsilon).
    \end{aligned}
\end{equation}
Here, the energy-dependent coefficients $A$ and $B$ are 
\begin{equation}
    \begin{aligned}
        &A(\epsilon)=\frac{4\tau_\text{im}}{L^2}\left[F_R(\epsilon)-F_L(\epsilon)\right]\frac{\frac{eV}{2}\sinh\frac{\epsilon}{T_0}-\epsilon\sinh\frac{eV}{2T_0}}{\cosh\frac{\epsilon}{T_0}-\cosh\frac{eV}{2T_0}}, \\
        &B(\epsilon)=\frac{4}{v_F^2\tau_\text{im}}\sinh\frac{eV}{2T_0}\frac{\frac{eV}{2}\sinh\frac{\epsilon}{T_0}-\epsilon\sinh\frac{eV}{2T_0}}{\cosh\frac{2\epsilon}{T_0}-\cosh\frac{eV}{T_0}}.
    \end{aligned}
\end{equation}

The total scattering time needed for evaluating the current and the noise power in this regime is approximately given by  
\begin{equation}
    \tau(x,\epsilon)\approx\tau_\text{im}\left[1-\tau_\text{im}2g^2(\epsilon+|e|Ex)\coth\frac{\epsilon+|e|Ex}{2T_0}\right],\label{eq:tauperturb}
\end{equation}
where the second term is the perturbative MFL contribution.
Note that this expression is valid only for $g^2\ll1$; otherwise, the scattering time in Eq.\eqref{eq:tauperturb} becomes negative, and the perturbative distribution function $F_s(x,\epsilon)$ in Eq.\eqref{eq:Fperturb} violates the Pauli principle (exceeds the bound between $-1$ to $1$) in certain parameter regimes, leading to unphysical results. In our numerical evaluation, we keep $g^2$ small enough such that these spurious cases do not arise.  

In Fig.\ref{fig:FanoC} we show the results for the Fano factor calculated based on the perturbation outlined above up to order $\mathcal{O}(g^2)$ (dropping $\mathcal{O}(g^4)$ terms). 
The $g^2=0$ data is the familiar result of Nagaev, $\mathcal{F}=\frac{1}{3}(\coth \frac{eV}{2T}-\frac{2T}{eV})$, which saturates to $1/3$ at $T_0=0$. Once a small MFL scattering is added, the Fano factor deceases with increasing MFL coupling strength, but the Fano factor at $T_0=0$ in our analysis depends on the strength of the MFL coupling and is not a universal number. 

\section{Conclusion and Discussion } 
\label{sec:discussion_and_conclusion}

In this paper, we investigate the intimate relation between the linear-in-$T$ resistivity and the vanishing shot noise of a strange metal, inspired by the recent experiment on YRS. We consider a phenomenological marginal Fermi liquid model, which describes fermions coupled to a collective boson mode with $\Omega/T$ scaling spectrum. We also include a disorder potential that becomes relevant at the lowest temperatures.  Our approach is built on the non-equilibrium quantum field theory, in which the quantum Boltzmann equation is viewed as the saddle point equation, and fluctuations around the saddle point lead to current noise. We have demonstrated that this approach, in the simplest case with only impurity scattering (i.e. a conventional dirty metal), coincides with the Boltzmann-Langevin approach.
With several approximations and simplifications stated in Sec.\ref{sec:shot_noise_in_a_pure_mfl}, we have derived and solved the Boltzmann equation in a more generic setting, and obtained the current noise for the three cases: 1) with only MFL scattering, 2) with large MFL scattering and small impurity scattering and 3) with large impurity scattering and small MFL scattering. 

Our main result is given by Eq.\eqref{eq:noisepower2} and \eqref{eq:MFLshot}: the total current noise power is a constant if the system is dominated by the MFL scattering, and therefore the shot noise vanishes. Adding impurity scattering perturbatively results in a finite but small temperature-dependent shot noise, as shown in the plot of Fano factor in Fig.\ref{fig:FanoB}(b). If instead the system is dominated by the impurity scattering and the MFL scattering is treated as a perturbation, as is more relevant in the $T\to0$ limit, the results for shot noise are shown in Fig.\ref{fig:FanoC}. In the limit when the MFL scattering is completely switched off, the shot noise reduces to Nagaev's result and the Fano factor saturates to $1/3$ as $T\to0$. Introducing MFL scattering tends to reduce the Fano factor, but the amount of reduction depends on the strength of MFL coupling and is thus not universal. 

We emphasize that the vanishing of shot noise is a unique feature of MFL scattering. Had we adopted another type of boson that leads to a more violent non-Fermi liquid behavior from the boson-fermion coupling, we would not obtain vanishing shot noise within the same theoretical treatment. Of course, that would also likely spoil the linear-in-$T$ resistivity behavior. This perhaps unveils the fact that both linear-in-$T$ resistivity and vanishing shot noise are two intimately related unique features of the strange metal phase. 

Lastly, we note that our treatment of the Boltzmann equation relies on the on-shell approximation in Eq.\eqref{eq:onshell}. A more rigorous yet rather involved way is to keep the full non-Fermi liquid spectral function in the collision integral and all renormalization effects from the singular fermion self energy. Sorting out the contributions from these terms that have been neglected here, to our knowledge, is still an open problem for a general non-Fermi liquid.  
It is also, in principle, straightforward to go beyond the accuracy of Boltzmann-Langevin and include higher order corrections to the noise power within the field-theoretic framework. We have outlined the derivation of higher order terms in Appendix \ref{sub:current_noise}, A detailed discussion about these corrections is left for future studies.

\begin{acknowledgments}
    We thank Alex Kamenev, Chandra Varma and Pavel Nosov for useful discussions. Y.-M.W. acknowledges the Gordon and Betty Moore Foundation’s EPiQS Initiative through GBMF8686 for support. J.J.Y.~was supported in part by the National Science Foundation Graduate Research Fellowship under Grant No.~DGE-1656518. Y.-M. W., J.J.Y. and S.R.  are supported in part by the US Department of Energy, Office of Basic Energy Sciences, Division of Materials Sciences and Engineering, under Contract No. DE-AC02-76SF00515. 
\end{acknowledgments}


\appendix
\section{MFL self energy at equilibrium}
\label{sec:MFLselfenergy}
Here we demonstrate the calculation of MFL self energy in the absence of external fields, by using both the equilibrium Matsubara frequency formalism.
The one-loop self energy in the Matsubara frequency formalism is given by 
\begin{equation}
    \begin{aligned}
        &\Sigma(\bk,i\epsilon_n)=g_0^2T\sum_m\int\frac{d^2\bp}{(2\pi)^2}\frac{D(\bp-\bk,i\epsilon_m-i\epsilon_n)}{i\epsilon_m-\xi_{\bp}}\\
        &=g_0^2T\sum_m\int\frac{d^2\bp}{(2\pi)^2}\int_{-\infty}^{+\infty} d\Omega\frac{-B(\Omega)}{\Omega-i\epsilon_m+i\epsilon_n}\frac{1}{i\epsilon_m-\xi_{\bp}}
    \end{aligned}
\end{equation}
where $B(\Omega)$ is the boson spectral function given in Eq.\eqref{eq:MFLspectral}. After performing the frequency summation, we can replace $i\epsilon_n$ with $\omega+i0^+$ to obtain the retarded self energy, 
\begin{equation}
    \Sigma^R(\bk,\omega)=\frac{g_0^2\nu_0}{\pi\omega_D}\int_{-\infty}^{+\infty}d\xi d\Omega \tanh\frac{\Omega}{2T}\frac{n_F(\xi)+n_B(\Omega)}{\xi-\Omega-\omega-i0^+}.
\end{equation}
The imaginary part of $\Sigma^R(\bk,\omega)$ is therefore easily obtained as 
\begin{equation}
    \begin{aligned}
        \text{Im}&\Sigma^R(\bk,\omega)=g^2\int_{-\infty}^{+\infty}d\Omega\tanh\frac{\Omega}{2T}\left[n_F(\omega+\Omega)+n_B(\Omega)\right]\\
        &=g^2\int_{-\infty}^{+\infty}d\Omega\tanh\frac{\Omega}{2T}\left[\coth\frac{\Omega}{2T}-\tanh\frac{\omega+\Omega}{2T}\right]\\
        &=g^2\omega \coth\frac{\omega}{2T}
    \end{aligned}\label{eq:ImSigma}
\end{equation}
where we have defined the dimensionless coupling constant $g^2=g_0^2\nu_0/(\pi\omega_D)$. If we idenfity $\text{Im}\Sigma^R(\bk,\omega)$ as $\frac{1}{2\tau_\text{MFL}}$, then the equilibrium MFL lifetime is 
\begin{equation}
    \tau_{\text{MFL}}=\frac{1}{2g^2}\frac{1}{\omega}\tanh\frac{\omega}{2T}
\end{equation}
which is consistent with the non-equilbirum MFL transport time given in Eq.\eqref{eq:tauMFL}.

\section{General formalism of kinetic theory}\label{sec:appB}
Our starting point is
a coupled boson-fermion problem, expressed in Keldysh fields, namely,
\begin{equation}
     \begin{aligned}
         S_\text{f} &= \int d\eta \bar{\Psi}(\eta)\hat{G}_0^{-1}\Psi(\eta),\\
         S_\text{b} &= \int d\eta d\eta' \Phi(\eta)\hat{D}^{-1}(\eta,\eta')\Phi(\eta'),\\
         S_\text{int} &= g_0\int d\eta \bar{\psi}_a(\eta)\gamma_{ab}^\alpha\psi_b(\eta)\phi^\alpha(\eta).
     \end{aligned}\label{eq:action1}
 \end{equation} 
 Here $\eta=(\br,t)$, $a,b=1,2$ and $\alpha=q,c$ and the summation over repeated indices is assumed. The vertex matrices $\gamma^\alpha$ are $\gamma^c=\sigma_0$, the $2\times2$ identity matrix, and $\gamma^q=\sigma_1$ is the first Pauli matrix, and $\Psi=(\psi_1, \psi_2)^T$ and $\Phi=(\phi^c, \phi^q)^T$ are the two-component Keldysh fields for fermions and bosons respectively. The fermionic Green's function (both bare and fully dressed) has the structure
 \begin{equation}
     \hat{G}=\begin{pmatrix}
         G^R & G^K\\
         0 & G^A
     \end{pmatrix}, ~~\hat{G}^{-1}=\begin{pmatrix}
         [G^R]^{-1} & [G^{-1}]^K\\
         0 & [G^A]^{-1}
     \end{pmatrix},
 \end{equation}
 while the bosonic Green's function is
  \begin{equation}
     \hat{D}=\begin{pmatrix}
         D^K & D^R\\
         D^A & 0
     \end{pmatrix}, ~~\hat{D}^{-1}=\begin{pmatrix}
         0 & [D^A]^{-1}\\
         [D^R]^{-1} & [D^{-1}]^K
     \end{pmatrix},
 \end{equation}
 where the superscripts $R, A$ and $K$ stand for retarded, advanced and Keldysh components. For the fermion part, $G^K= G^R F- F G^A$ where $F$ is the generalized Fermi distribution function. And similarly $D^K= D^R F_b- F_b D^A$ with $F_b$ the generalized boson distribution function.  With this definition it's easy to see that $[G^{-1}]^K=-[G^R]^{-1}G^K[G^A]^{-1}=[G^R]^{-1}F-F[G^A]^{-1}$ and similarly $[D^{-1}]^K=-[D^R]^{-1}D^K[D^A]^{-1}=[D^R]^{-1}F_b-F_b[D^A]^{-1}$.
In equilibrium, the fermion and boson distribution funciton take the familiar forms
\begin{equation}
    \begin{aligned}
        F(\epsilon)&=1-2n_F(\epsilon)=\tanh\frac{\epsilon}{2T},\\
        F_b(\epsilon)&=1+2n_B(\epsilon)=\coth\frac{\epsilon}{2T}
    \end{aligned}
\end{equation}
Note that in the fermion sector, we start with the free fermion action, for which $\hat{G}_0$ is local in space. In the bosonic sector, however, we inherit the assumption that the Green's function $\hat{D}(\eta,\eta')$ already contains dynamically generated retardation effect. This is different from the free boson Green's function $\hat{D}_0$.

 From Eq.\eqref{eq:action1}, one can integrate out the bosons to obtain the fermion-only action
\begin{equation}
    \begin{aligned}
        S_\text{f}&=\int d\eta \bar{\psi}_a(\eta)[\hat{G}_0^{-1}]_{ab}\psi_b(\eta)\\
        &-\frac{g_0^2}{4}\int dx dx' \bar{\psi}_a(\eta)\gamma_{ab}^\alpha\psi_b(\eta)D_{\alpha\beta}(\eta,\eta')\bar{\psi}_c(\eta')\gamma^\beta_{cd}\psi_d(\eta').
    \end{aligned}
\end{equation}
We next introduce a fluctuating bilocal field $\tilde G_{ab}(\eta,\eta')=-i\psi_{a}(\eta)\bar{\psi}_b(\eta')$, and enforce this relation using a Lagrange multiplier field $\tilde\Sigma_{ab}(\eta,\eta')$. In practice, this is to insert the resolution of identity
\begin{equation}
    \begin{aligned}
        1\equiv\int \mathcal{D}[\tilde G,\tilde \Sigma]\exp&\left\{\int d\eta d\eta'\tilde \Sigma_{ab}(\eta,\eta')\right.\\
        &\left.\left[\tilde G_{ba}(\eta',\eta)+i\psi_b(\eta')\bar{\psi}_a(\eta)\right]\right\}.
    \end{aligned}
\end{equation}
The $\tilde G$-$\tilde \Sigma$ only action can be obtained after further integrating out the fermions,
\begin{equation}
    S[\tilde G,\tilde\Sigma]=-i\text{Tr}\ln(1-\tilde G_0\tilde\Sigma)-i\tilde G\tilde\Sigma  -\frac{g_0^2}{4}( \tilde G\gamma^\beta \tilde G \gamma^\alpha D_{\alpha\beta}).\label{eq:GSigma}
\end{equation}
Note in the last part, the first $\tilde G$ is $\tilde G(\eta,\eta')$, while the second is $\tilde G(\eta',\eta)$.
The Dyson equation and the consistent equation for the self energy are obtained by taking the saddle point of $S[\tilde G,\tilde\Sigma]$:
\begin{equation}
    \begin{aligned}
        &\frac{\delta S}{\delta\tilde\Sigma}=0 ~ \to ~ \hat G=(1-\hat G_0 \hat\Sigma)^{-1}\hat G_0=(\hat G_0^{-1}-\hat\Sigma)^{-1}\\
        &\frac{\delta S}{\delta\tilde G}=0 ~\to~ \hat\Sigma =i\frac{g_0^2}{2}\gamma^\beta\hat G\gamma^\alpha D_{\alpha\beta},
    \end{aligned}\label{eq:saddleeq}
\end{equation}
where $\hat G$ and $\hat \Sigma$ are the saddle point solutions of $\tilde G$ and $\tilde \Sigma$.
From the first equation, we can see that $\hat\Sigma$ has the same matrix structure as $\hat G^{-1}$, which we denote as 
\begin{equation}
    \hat\Sigma=\begin{pmatrix}
        \Sigma^R & \Sigma^K\\
        0 & \Sigma^A
    \end{pmatrix}.\label{eq:matrixSig}
\end{equation}
With this the equations for the Green's functions can be written as 
\begin{equation}
    \begin{aligned}
        [G^R]^{-1} &= [G^R_0]^{-1}- \Sigma^R,\\
        [G^A]^{-1} &= [G^A_0]^{-1}- \Sigma^A,\\
        F[G^A_0]^{-1}-[G^R_0]^{-1}F&=\Sigma^K-(\Sigma^R F - F\Sigma^A).
    \end{aligned}\label{eq:G1}
\end{equation}
In fact, for the free Green's functions we can set $G_0^R=G_0^A=G_0$, then the last equation in above can be rewritten as
\begin{equation}
    -[G_0^{-1},F]=\Sigma^K-(\Sigma^R F - F\Sigma^A).\label{eq:kinetic1}
\end{equation}
All the operators in the above equation should be treated as a matrix and the operator product is implemented as the matrix product. 

The equations for $\Sigma^R, \Sigma^A$ and $\Sigma^K$ are obtained from the second equation, which are explicitly 
\begin{equation}
    \begin{aligned}
        \Sigma^R(\eta_1,\eta_2)=i\frac{g_0^2}{2}&\left(G^R(\eta_1,\eta_2)D^K(\eta_2,\eta_1)\right.\\
        &\left.+G^K(\eta_1,\eta_2)D^A(\eta_2,\eta_1)\right)\\
        \Sigma^A(\eta_1,\eta_2)=i\frac{g_0^2}{2}&\left(G^A(\eta_1,\eta_2)D^K(\eta_2,\eta_1)\right.\\
        &\left.+G^K(\eta_1,\eta_2)D^R(\eta_2,\eta_1)\right)\\
        \Sigma^K(\eta_1,\eta_2)=i\frac{g_0^2}{2}&\left[G^K(\eta_1,\eta_2)D^K(\eta_2,\eta_1)\right.\\
        -\left(G^R(\eta_1,\eta_2)\right.-&\left.G^A(\eta_1,\eta_2)\right)\left(D^R(\eta_2,\eta_1)-D^A(\eta_2,\eta_1)\right)].
    \end{aligned}\label{eq:Sigma1}
\end{equation}
where in the last line the causality results in $G^R(\eta_1,\eta_2)D^R(\eta_2,\eta_1)=G^A(\eta_1,\eta_2)D^A(\eta_2,\eta_1)=0$.

When the system is exposed to some static electric field, the fermion density couples to the scalar field $V(\eta)\equiv V(\br)$. Passing this field to the Keldysh contour and assuming the potential is identical along the forward and backward path, the only relevant fields is the classical component $V^c(\eta)=\frac{1}{2}(V^+(\eta)+V^-(\eta))$. The quantum component $V^q(\eta)=\frac{1}{2}(V^+(\eta)-V^-(\eta))$ vanishes, although it can be kept nonzero in order to generate density correlation functions before setting to zero at the end of calculation. With this potential, and assuming the quadratic dispersion, the free fermion inverse Green's functions are given by
\begin{equation}
    [G^{R(A)}_0]^{-1}=i\partial_t-\frac{(-i\nabla_{\br})^2}{2m}-V^c(\br)\pm i0^+.\label{eq:2dispersion}
\end{equation}

We are now in a position to solve the coupled equations Eq.\eqref{eq:G1} and Eq.\eqref{eq:Sigma1} perturbatively. Eq.\eqref{eq:kinetic1} after Wigner transform becomes
\begin{equation}
    \begin{aligned}
        -i&\left[\partial_t+\bm{v}\cdot\nabla_{\br}-\left(\nabla_{\br}V^c(\br)\right)\cdot\nabla_{\bk}\right]F(\eta,k)\\
        =&\Sigma^K(\eta,k)-2i\text{Im}\Sigma^R(\eta,k)F(\eta,k)\\
        &-i\left(\nabla_{\br}\text{Re}\Sigma^R\cdot\nabla_{\bk}-\partial_t\text{Re}\Sigma^R\partial_\omega\right.\\
        &\left.-\nabla_{\bk}\text{Re}\Sigma^R\cdot\nabla_{\br}+\partial_\omega\text{Re}\Sigma^R\partial_t\right)F(\eta,k).
    \end{aligned}\label{eq:kinetic2}
\end{equation}
where we remind that $\eta=(\br,t)$ and $k=(\bk,\omega)$.
In a Fermi liquid system, one argues that the distribution function $F(\eta,k)$ can be approximated by $F(\br,t,\bk)$, with the frequency peaked at $\omega=\epsilon_{\bk}+V^r(\br)+\text{Re}\Sigma^R(\eta,k)$. This ``mass-shell'' argument depends on the existence of well-defined quasi-particles. Hence in our case of non-Fermi liquid, abandoning the $\omega$-dependence is not justified and we will keep the full $\omega$-dependence of $F$. We can introduce 
\begin{equation}
    \begin{aligned}
        Z^{-1}(\eta,k)&=1-\partial_\omega\text{Re}\Sigma^R(\eta,k)\\
        \bm{v}^*(\eta,k)&=\nabla_{\bk}[\epsilon_{\bk}+\text{Re}\Sigma^R(\eta,k)]\\
        \tilde{V}(\eta,k)&=V^c(\br)+\text{Re}\Sigma^R(\eta,k)
    \end{aligned}
\end{equation}
and rewrite Eq.\eqref{eq:kinetic2} as follows
\begin{equation}
    \left(Z^{-1}\partial_t+\bm{v}^*\cdot\nabla_{\br}-\nabla_{\br}\tilde{V}\cdot\nabla_{\bk}+\partial_t\text{Re}\Sigma^R\partial_\omega\right)F(\eta,k)=I_\text{col},\label{eq:kinetic2}
\end{equation}
and the right hand side is defined as the collision integral $I_\text{col}=i\Sigma^K(\eta,k)+2\text{Im}\Sigma^R(\eta,k)F(\eta,k)$, which is evaluated as 
\begin{widetext}
\begin{equation}
    \begin{aligned}
            I_\text{col}=-\frac{g_0^2}{2}&\left\{\int_p[G^R(\eta,p)-G^A(\eta,p)][F(\eta,p)-F(\eta,k)]D^K(\eta,p-k)\right.\\
            &-\left.\int_p[G^R(\eta,p)-G^A(\eta,p)][1-F(\eta,p)F(\eta,k)][D^R(\eta,p-k)-D^A(\eta,p-k)]\right\}
    \end{aligned}\label{eq:Icol1}
\end{equation}
If we define
\begin{equation}
    G^R(\eta,k)-G^A(\eta,k)=-i2\pi A(\eta,k), ~~ D^R(\eta,q)-D^A(\eta,q)=-i2\pi B(\eta,q) \label{eq:bosonD3}
\end{equation}
where $A$ and $B$ are the fermion and boson spectral functions respectively, we can write the above equation as
\begin{equation}
    \begin{aligned}
            I_\text{col}(\eta,k)={2}\pi^2g_0^2&\int_pA(\eta,p)\left\{[F(\eta,p)-F(\eta,k)]F_b(\eta,p-k)-[1-F(\eta,p)F(\eta,k)]\right\}B(\eta,p-k)\\
    \end{aligned}\label{eq:generalcollision}
\end{equation}
\end{widetext}
Below we discuss, based on a few examples, about the explicit form of the collision integral in Eq.\eqref{eq:Icol1} or Eq.\eqref{eq:generalcollision}. Like in the main text, we will be seeking for the steady state solution of the distribution functions, which are thus $t$-independent. We also assume, in the presence of the electric field pointing towards $x$-direction, the distribution functions therefore depend only on $x$ so the notation $\eta$ is reduced to $x$ and $F(\eta,k)\to F(x,k)$.

\subsection{impurity scattering}
Here we consider a very simple model: free electrons moving in an electric field and getting scattered by impurities. The impurity is described by a purely classical disorder potential $V_{dis}(\br)$, which we assume to have zero mean and the following correlation
\begin{equation}
    \overline{V_{dis}(\br)V_{dis}(\br')}=g(\br-\br').
\end{equation}
This is obtained from assuming the disorder distribution function to be 
\begin{equation}
    P[V]=\exp\left[-\frac{1}{2}\int d\br d\br' V_{dis}(\br)g^{-1}(\br-\br') V_{dis}(\br')\right].\label{eq:disV}
\end{equation}
Since we only consider the classical component of $V_{dis}$, this correlator plays the role of $D^K$ in Eq.\eqref{eq:Icol1}, and other components $D^{R(A)}$ are absent. As a result, the collision integral from this model is 
\begin{equation}
    I_\text{im}(x,\bk)=2\pi\int_{\bp}g(\bk-\bp)\left[F(x,\bp)-F(x,\bk)\right]\delta(\epsilon_{\bp}-\epsilon_{\bk}).
\end{equation}
Here we have used the on-shell approximation for $F(x,k)$, meaning $F(x,k)=F(x,\bk)\delta(\omega-\epsilon_{\bk})$. The delta-function is then from the fact that the impurity scattering is elastic which does not transfer energy. 

Next, we make use of Eq.\eqref{eq:partialwave} and write $F(x,\bk)$ in terms of $F_s(x,\epsilon)$ and $F_p(x,\epsilon)$, and define
\begin{equation}
    I_{\text{im},s}=\int_{-\pi}^{\pi}\frac{d\theta_{\bk}}{2\pi}I_\text{im}(x,\bk), ~I_{\text{im},p}=\int_{-\pi}^{\pi}\frac{d\theta_{\bk}}{2\pi}I_\text{im}(x,\bk)\cos\theta_{\bk}.
\end{equation}
A straightforward calculation then shows,
\begin{equation}
    I_{\text{im},s}=0,~~ I_{\text{im},p}=-\frac{1}{\tau_\text{im}} F_p(x,\epsilon),\label{eq:sptauIM}
\end{equation}
where the impurity scattering time is defined as 
\begin{equation}
   \frac{1}{\tau_\text{im}}= \nu_0\int_{-\pi}^\pi d\theta g(\theta)(1-\cos\theta).
\end{equation}

\subsection{electron-electron scattering}
The collision integral from electron-electron scattering vanishes if we use the bare electron interaction in Eq.\eqref{eq:Icol1}. The familar expression for the electron-electron collision integral is obtained by using the RPA dressed interaction in Eq.\eqref{eq:Icol1}. Let's assume the bare, short range electron-electron interaction is $U$, and the boson mode is introduced via the Hubbard-Stratonovich transformation. We use the convention that the boson-fermion coupling constant is $g_0=1$ such that the bare boson kernel is $\hat D_0^{-1}=U^{-1}\sigma_1$. The RPA boson Green's funciton is then 
\begin{equation}
    \hat D^{-1}=\hat D_0^{-1}+\hat\Pi_0,\label{eq:RPA}
\end{equation}
where $\hat\Pi_0$ is the one-loop fermionic bubble. Since $\hat\Pi_0$ plays the same role as the boson self energy, like in Eq.\eqref{eq:matrixSig} we can write it as
\begin{equation}
    \hat\Pi_0=\begin{pmatrix}
        0 & \Pi_0^A\\
        \Pi_0^R & \Pi_0^K
    \end{pmatrix}
\end{equation}
where the regarded, advanced and Keldysh bubbles are (we have suppressed the spatial dependence)
\begin{equation}
    \begin{aligned}
        &\Pi_0^{R(A)}(\bq,\omega)=-\frac{1}{2}\int\frac{d^2\bk}{(2\pi)^2}\frac{F(\bk+\bq)-F(\bk)}{\omega+\epsilon_{\bk}-\epsilon_{\bk+\bq}\pm i0^+}, \\
        &\Pi_0^{K}(\bq,\omega)=-i\pi\int\frac{d^2\bk}{(2\pi)^2}[F(\bk)F(\bk+\bq)-1]\delta(\omega+\epsilon_{\bk}-\epsilon_{\bk+\bq}).
    \end{aligned}
\end{equation}

Inverting Eq.\eqref{eq:RPA}, one obtains for the retarded, advanced and Keldysh Green's functions
\begin{equation}
    \begin{aligned}
        D^{R(A)}&=(U^{-1}+\Pi_0^{R(A)})^{-1}\\
        D^{K}&=-D^R\Pi_0^K D^A.
    \end{aligned}
\end{equation}
These relations, after Wigner transform, lead to 
\begin{widetext}
\begin{equation}
   \begin{aligned}
        D^K(q)&=i\pi|D^R(q)|^2\int_{\bk'}[F(\bk')F(\bk'+\bq)-1]\delta(\omega+\epsilon_{\bk'}-\epsilon_{\bk'+\bq}),\\
        D^R(q)-D^A(q)&=-i\pi|D^R(q)|^2\int_{\bk'}[F(\bk'+\bq)-F(\bk')]\delta(\omega+\epsilon_{\bk'}-\epsilon_{\bk'+\bq}).
   \end{aligned}
\end{equation}    
Substituting this into Eq.\eqref{eq:Icol1}, and make use of the on-shell approximation for the distribution functions, as well as Fermi liquid like fermion spectral function, we obtain 
\begin{equation}
    \begin{aligned}
        I_{ee}=-\frac{\pi}{2}\int_{\bk',\bq}|D^R(\bq,\epsilon_{\bk'+\bq}-\epsilon_{\bk'})|^2&\left\{[F(\bk+\bq)-F(\bk)][F(\bk')F(\bk'+\bq)-1]-[F(\bk+\bq)F(\bk)-1][F(\bk'+\bq)-F(\bk')]\right\}\\
        &\times\delta(\epsilon_{\bk}-\epsilon_{\bk+\bq}-\epsilon_{\bk'}+\epsilon_{\bk'+\bq}).
    \end{aligned}
\end{equation}
Using the relation $F=1-2n_F$, it is easy to see that the collision integral can be rewritten as 
\begin{equation}
    \begin{aligned}
        I_{ee}=-4\pi\int_{\bk',\bq}|D^R(\bq,\epsilon_{\bk'+\bq}-\epsilon_{\bk'})|^2&\left\{n_F(\bk')n_F(\bk+\bq)[1-n_F(\bk)][1-n_F(\bk'+\bq)]\right.\\
        &-n_F(\bk)n_F(\bk'+\bq)[1-n_F(\bk')][1-n_F(\bk+\bq)]\}\times\delta(\epsilon_{\bk}-\epsilon_{\bk+\bq}-\epsilon_{\bk'}+\epsilon_{\bk'+\bq}).
    \end{aligned}
\end{equation}
which is the well known formula for the e-e scattering.
\end{widetext}

\subsection{MFL scattering}
As the last example, we discuss how the MFL scattering due to the coupling to the boson mode in Eq.\eqref{eq:MFLspectral} gives rise to a collision integral that is very similar to the impurity scattering case. As stated in the main text, the key assumption we adopt is that the bosons are in local equilibrium, and thus their Green's functions obey the local FDT,
\begin{equation}
    D^K(\bq,\Omega)=\coth\frac{\Omega}{2T(x)}\left(D^R(\bq,\Omega)-D^A(\bq,\Omega)\right).\label{eq:equiboson}
\end{equation}
This means the boson distribution funciton $F_b$ in Eq.\eqref{eq:generalcollision} simply becomes
\begin{equation}
    F_b(x,\Omega)=\coth\frac{\Omega}{2T(x)},
\end{equation}
and inserting the boson spectral function in Eq.\eqref{eq:MFLspectral} the collision integral within the on-shell approximation is
\begin{equation}
    \begin{aligned}
        I_\text{MFL}={{2}\pi g^2}&\int \frac{d\epsilon_{\bp} d\theta_{\bp}}{(2\pi)^2}\left\{\left[F(\br,\bp)-F(\br,\bk)\right]\right.\\
       &-\left.\left[1-F(\br,\bp)F(\br,\bk)\right]\tanh\frac{\epsilon_{\bp}-\epsilon_{\bk}}{2T(x)}\right\},
    \end{aligned}
\end{equation}
where the dimensionless coupling constant $g^2$ is the same as that in Eq.\eqref{eq:ImSigma}. Making use of the decomposition in Eq.\eqref{eq:partialwave}, it is easy to obtain the $s$- and $p$-wave components of $I_\text{MFL}$, which are given in Eq.\eqref{eq:IMFLsp}. Similar to
Eq.\eqref{eq:sptauIM}, after we adopt the ansatz in Eq.\eqref{eq:ansatz2}  we have $I_{\text{MFL},s}=0$ and  $I_{\text{MFL},p}$ is given by Eq.\eqref{eq:IMFLsp}.

\section{Field-theoretic description of current and noise} 
\label{sec:field_theoretic_description_of_current_noise}
To study transport properties of the model discussed above, we first obtain the current operator. For simplicity, we keep assuming the quadratic dispersion in Eq.\eqref{eq:2dispersion}. In this case, introducing coupling to external gauge field is simply replacing $-i\nabla_{\br}$ with $-i\nabla_{\br}-e\bm{A}$. If we focus on the paramagnetic contribution to the current, it amounts to changing $G_0^{-1}$ to $G_0^{-1}+e\bm{v}\cdot\bm{A}$ where $\hat{\bm{v}}=-i\nabla_{\br}/m$. To get the current operator and its correlations, we consider the quantum component of this vector potential, which couples to the fermion fields as
\begin{equation}
    S_A=\int dx \bar{\psi}_a(x)\gamma_{ab}^q\psi_{b}(x)e\bm{v}\cdot\bm{A}^q(x).
\end{equation}
This is nothing but the vector potential couples linearly to the fluctuating field $\tilde G$.
After integrating out the fermions, we obtain
\begin{equation}
   \begin{aligned}
        S_A[\tilde G,\tilde\Sigma]&=S[\tilde G,\tilde\Sigma]-i\text{Tr}\ln\left(1+(G_0^{-1}-\tilde\Sigma)^{-1}\gamma^qe\bm{v}\cdot\bm{A}^q\right)\\
        &\approx S[\tilde G,\tilde\Sigma]-i\text{Tr}\ln\left(1+\tilde G\gamma^qe\bm{v}\cdot\bm{A}^q\right)\\
        &= S[\tilde G,\tilde\Sigma]-i\text{Tr}\left(\tilde G\gamma^qe\bm{v}\cdot\bm{A}^q\right)\\
        &~~~+\frac{i}{2}\text{Tr}\left(\tilde G\gamma^qe\bm{v}\cdot\bm{A}^q\tilde G\gamma^qe\bm{v}\cdot\bm{A}^q\right)
   \end{aligned}
\end{equation}
where $S[\tilde G,\tilde\Sigma]$ is given in Eq.\eqref{eq:GSigma}, and in the second line we have used the saddle point condition that leads to Dyson's equation. In general $\tilde G$ and $\tilde\Sigma$ fluctuate independently, However, once we demand $\delta S[\tilde G,\tilde\Sigma]/\delta\tilde\Sigma=0$, which leads to the Dyson's equation, the fluctuations of $\tilde G$ and $\tilde\Sigma$ are pinned together. As a result, only fluctuations of $\tilde G$ need to be considered in the following analysis. From the above equation we can identify the current operators as 
\begin{equation}
    \bm{J}(x):=\frac{1}{2}\left.\frac{\delta S_A}{\delta \bm{A}^q(x)}\right|_{\bm{A}^q=0}=-i\frac{e}{2}\text{Tr}(\tilde{G}(x,x')\gamma^q\bm{v})|_{x=x'}.\label{eq:current1}
\end{equation}
Since $\tilde G$ is a fluctuating field, so is the current operator.
Near the saddle point, we can write $\tilde{G}=\hat{G}+\delta G$ where $\hat{G}$ is the saddle point Green's function, and $\delta G$ accounts for fluctuations. It is this fluctuation $\delta{G}$ that leads to the noises of the current. 

\subsection{Fluctuations above the saddle point} 
\label{sub:fluctuations_above_the_saddle_point}

To study the noise effect of the current, we need to rewrite the $\tilde G$-$\tilde \Sigma$ action in terms of the fluctuation fields. We adopt the convention by Kamenev\cite{kamenev2023field} that 
\begin{equation}
     \begin{aligned}
      &   \tilde G= \mathcal{U}^{-1}\circ\begin{pmatrix}
         G^R & 0\\
         i\delta_G & G^A
     \end{pmatrix}\circ \mathcal{U},\\
     & \tilde\Sigma=\mathcal{U}^{-1}\circ\begin{pmatrix}
         \Sigma^R & \Sigma^K-\Sigma^R\circ F+ F\circ\Sigma^A\\
         i\delta_\Sigma & \Sigma^A
     \end{pmatrix}\circ \mathcal{U}
     \end{aligned}
 \end{equation} 
where 
\begin{equation}
    \mathcal{U}=\begin{pmatrix}
        1 & F\\
        0 & 1
    \end{pmatrix}, ~~\mathcal{U}^{-1}=\begin{pmatrix}
        1 & -F\\
        0 & 1
    \end{pmatrix}.
\end{equation}
So we have, for example, 
\begin{equation}
    \begin{aligned}
       \tilde G &= \begin{pmatrix}
        G^R-iF\circ\delta_G & G^R\circ F- F \circ G^A -iF\circ\delta_G\circ F\\
        i\delta_G & i\delta_G\circ F+G^A
    \end{pmatrix}\\
    &=\hat G+i\begin{pmatrix}
        -F\circ\delta_G &  -F\circ\delta_G\circ F\\
        \delta_G & \delta_G\circ F
    \end{pmatrix}.
    \end{aligned}\label{eq:tildeGdeltaG}
\end{equation}

With this definition, Eq.\eqref{eq:GSigma} can be rewritten in terms of $\delta_G$ and $\delta_\Sigma$ fields, and the functional integral is now over these new fluctuating fields. Expanding the new action to Gaussian level, $S[\delta_G,\delta_\Sigma]$ contains two part. The first part contains linear in $\delta_G$ and $\delta_\Sigma$ terms. Requiring $\frac{\delta S}{\delta[\delta_{G,\Sigma}]}|_{\delta_{G,\Sigma}=0}=0$ reproduces the above equations in Eq.\eqref{eq:saddleeq}. So these linear terms vanish at the saddle point. The second part, which is relevant for fluctuations, is quadratic in $\delta_G$ (note there are no terms containing $\delta_\Sigma^2$ or $\delta_G\delta_\Sigma$ under this convention). After Wigner transform it is given by 
\begin{equation}
    iS[\delta_G]=\frac{g^2}{4}\int dx dk dp\delta_G(x,k)K(x,k,p)\delta_G(x,p)\label{eq:iSdeltaG}
\end{equation}
where
\begin{widetext}
\begin{equation}
    K(x,k,p)=i[F(x,k)-F(x,p)]\left\{[F(x,p)-F(x,k)]D^K(x,p-k)+[F(x,k)F(x,p)-1](D^R(x,p-k)-D^A(x,p-k))\right\}.
\end{equation}
In the case of MFL, the bosonic Green's functions are given by Eqs.\eqref{eq:MFLspectral},\eqref{eq:bosonD3} and \eqref{eq:equiboson} and thus
\begin{equation}
    K(x,k,p)=2\nu_0[F(x,k)-F(x,p)]\left\{[F(x,p)-F(x,k)]+[F(x,k)F(x,p)-1]\tanh\frac{\omega'-\omega}{2T}\right\}.\label{eq:Kernel_general}
\end{equation}
Note if $F$ is the equilibrium distribution function, this kernel $K$ vanishes identically. 
The non-equilibrium distribution function $F(x,k)$ is solved from the saddle point equations, in particular, Eq.\eqref{eq:generalcollision}. For this Gaussian fluctuations, the average $\braket{\delta_G}=0$, and the correlation $\braket{\delta_G(x,k)\delta_G(x',p)}$ is determined by the inverse of $K$. Since $K$ vanishes at equilibrium, $\delta_{G}$-correlation contains a singular piece which requires care in practice. 

The Gaussian fluctuations of $\delta_G$ can  be further decomposed to linear level by introducing a noise field $\xi$ via Hubbard-Stratonovich transformation, resulting in a term like $\delta_G\xi$. This Gaussian noise field has zero mean and its correlation is given by $K(x,k,p)$. And it enters the saddle point equation, making Eq.\eqref{eq:generalcollision} into the Boltzmann-Langevin (BL) equation. One could in principle solve the BL equation to obtain the current fluctuations. 

Here we avoid introducing the noise field. Instead, we express the current in terms of the fluctuating field $\tilde G$ as in Eq.\eqref{eq:current1}. It can be also written in terms of $\delta_G$, 
\begin{equation}
    \begin{aligned}
        \bm{J}(x)&=-i\frac{e}{2}\lim_{x\to x'}\left(G^R\circ F-F\circ G^A-iF\circ\delta_G\circ F+i\delta_G\right)\bm{v}\\
        &=-\pi e\int_k A(x,k)F(x,k)\bm{v}_{\bm{k}}+\frac{e}{2}\int_k[1-F^2(x,k)]\delta_G(x,k)\bm{v}_{\bm{k}}.
    \end{aligned}
\end{equation}
    
\end{widetext}
From this, it is easy to see that the average current is
\begin{equation}
   \braket{\bm{J}(x)}:=-\left.\frac{i}{2}\frac{\delta\ln Z[\bm{A}]}{\delta\bm{A}^q(x)}\right|_{\bm{A}^q\to0}=-\pi e\int_k A(x,k)F(x,k)\bm{v}_{\bm{k}}
\end{equation}
due to the fact that $\braket{\delta_G}=0$.  For a Fermi liquid in a steady (DC) external field, $A(x,k)=\delta(\omega-\epsilon_{\bk}-V(x))$ so the above expression reduces to
\begin{equation}
    \braket{\bm{J}(x)}=-\frac{e}{2}\sum_{\bk}F(x,\bk)\bm{v}_{\bm{k}},\label{eq:averageJ}
\end{equation}
which is the well known result for Fermi liquid. From this it is clear that to induce a finite current, the inversion symmetry must be broken, otherwise the momentum summation leads to zero. 
For non-Fermi liquid, however, we have to keep the full frequency dependence of $A(x,k)$ and $F(x,k)$.

\begin{figure}
     \includegraphics[width=8cm]{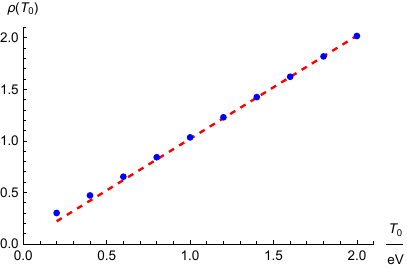}
     \caption{Resistivity as a function of $T_0$, calculated from Eq.\eqref{eq:MFLAverageCurrent}. Here $\rho$ is in units of $\frac{m4\pi^2g^2\nu_0^2T_0}{7\zeta(3)ne^2}$. The red dashed line indicates the linear-in-$T$ behavior.  At larger $T_0$, $\rho$ follows this line. At lower $T_0$, $\rho$ deviates from the linear-in-$T$ behavior.  }\label{fig:linear-in-T}
 \end{figure}

As an example, let's try to calculate the average current density for the MFL scattering. Using Eq.\eqref{eq:averageJ}, we have
 \begin{equation}
     \begin{aligned}
         \braket{\bm{J}(x)}&=-\frac{e\nu_0}{2}\int d\epsilon \int \frac{d\theta}{2\pi}(F_s+\cos\theta F_p)v_F\cos\theta\\
         &=-\frac{e\nu_0v_F}{4}\int d\epsilon F_p(x,\epsilon)\\
         &=\frac{e\nu_0v_F^2}{4}\int d\epsilon \tau_{ee}(x,\epsilon)\partial_xF_s(x,\epsilon).
     \end{aligned}
 \end{equation}
 With the assumption of $F_s(x,\epsilon)$
 the average current is given by
 \begin{equation}
    \begin{aligned}
         \braket{\bm{J}(x)}=\frac{e\nu_0v_F^2}{4}&\int d\epsilon \tau(x,\epsilon)\text{sech}^2\frac{\epsilon-\mu(x)}{2T(x)}\times\\
         &\left[\frac{eE}{2T(x)}-\frac{\epsilon-\mu(x)}{2T^2(x)}\frac{\partial T(x)}{\partial x}\right].
    \end{aligned}
 \end{equation}
 For fixed $x$, the last term in the bracket vanishes after integrating over $\epsilon$. If $\tau_{ee}$ is a constant, the above result reduces to $\braket{\bm{J}}=\sigma E$ where $\sigma=ne^2\tau/m$. With MFL scattering time,  we obtain
 \begin{equation}
     \braket{\bm{J}}=\frac{7\zeta(3)e^2v_F^2\nu_0D_0}{4\pi^2}\frac{E}{\bar{T}}\equiv\frac{ne^2\tau(\bar T)}{m}E, 
     \label{eq:MFLAverageCurrent}
 \end{equation}
 where we have defined $\bar{T}$ as some average temperature,
 \begin{equation}
     {\bar T}^{-1}=\frac{1}{L}\int_{-L/2}^{L/2}dx\frac{1}{T(x)}.
 \end{equation}
 Note that this $\bar T$ is different from $T_0$ and the deviation is set by the electric field $E$. Keeping such deviations means we go beyond the linear response regime.  
 In Fig.\ref{fig:linear-in-T}
 we show the properly normalized resistivity as a function of temperature. We see that as long as the temperature is not vanishingly small compare to the applied voltage, $\rho(T_0)$ shows linear dependence on $T_0$, the environment temperature. 

\subsection{Current noise} 
\label{sub:current_noise}
We define the current fluctuation as the spatially averaged one, which is given by
\begin{equation}
    \delta I(t)=\frac{1}{\Omega}\int d\br \delta\bm{J}(\br,t)\cdot \Delta S
\end{equation}
where $\Omega$ is the system volume and $\Delta S$ is the cross section area. In our case of a two dimensional system, $\Delta S=W$ is the width of the sample and $\Omega=LW$. 
The noise power at temperature $T$ subject to some external electric voltage $V$ is then defined by
\begin{equation}
    \begin{aligned}
        \mathcal{S}(T,V)&=\int dt \braket{\delta I(t)\delta I(0)+\delta I(0)\delta I(t)}\\
        &=-\frac{W^2}{2\Omega^2}\int d\br d\br' dt\left.\frac{\delta^2 \ln Z[\bm{A}]}{\delta\bm{A}^q(x)\delta\bm{A}^q(x')}\right|_{\bm{A}^q\to0}
    \end{aligned}
\end{equation}
\begin{widetext}
Using the expression for $Z[\bm{A}]$ introduced above, it is easy to see that
\begin{equation}
    \begin{aligned}
        \mathcal{S}(T,V)&=\frac{e^2W^2}{2\Omega^2}\int d\br d\br' dt\left\{\braket{\text{Tr}\left[\tilde G(x,x')\gamma^q \bm{v}\tilde G(x',x)\gamma^q \bm{v}\right]}-\braket{\text{Tr}\left[\tilde G(x,x)\gamma^q\bm{v}\right]\text{Tr}\left[\tilde G(x',x')\gamma^q\bm{v}\right]}+\braket{\text{Tr}\left[\tilde G(x,x)\gamma^q\bm{v}\right]}^2\right\}.\label{eq:S1}
    \end{aligned}
\end{equation}
Making use of Eq.\eqref{eq:current1}, we can also rewrite the above expression as 
\begin{equation}
    \begin{aligned}
        \mathcal{S}(T,V)&=\frac{W^2}{4\Omega^2}\int d\br d\br' dt\left\{e^2\braket{\text{Tr}\left[\tilde G(x,x')\gamma^q \bm{v}\tilde G(x',x)\gamma^q \bm{v}\right]}+4\braket{\bm{J}(x)\bm{J}(x')}-4\braket{\bm{J}}^2\right\}.
    \end{aligned}
\end{equation}
The first term in Eq.\eqref{eq:S1} (contains both JN noise and shot noise as we will see below) can be evaluated as 
\begin{equation}
    \begin{aligned}
        &\int d\br d\br' dt\braket{\text{Tr}\left[\tilde G(x,x')\gamma^q \bm{v}\tilde G(x',x)\gamma^q \bm{v}\right]}\\
        =&\int d\br\sum_k \bm{v}_{\bk}^2[F^2(x,k)-1][G^R(x,k)-G^A(x,k)]^2-\int d\br\sum_k \bm{v}_{\bk}^2[F^2(x,k)-1]^2\braket{\delta_G(x,k)\delta_G(x,k)}.\label{eq:S2}
    \end{aligned}
\end{equation}
Here the integration over $t$ is equivalent to integrate over the relative time $t-t'$ and thus restrict the two frequencies in $F(x,k)$ and $F(x',p)$ to be the same. Similar facts also applies to the space integration. In the last line $d\br$ should be understood as the center of mass coordinate. 

The last two terms in Eq.\eqref{eq:S1} lead to
\begin{equation}
    \begin{aligned}
        &\int d\br d\br' dt\left\{-\braket{\text{Tr}\left[\tilde G(x,x)\gamma^q\bm{v}\right]\text{Tr}\left[\tilde G(x',x')\gamma^q\bm{v}\right]}+\braket{\text{Tr}\left[\tilde G(x,x)\gamma^q\bm{v}\right]}^2\right\}\\
        =&\int d\br d\br' dt \sum_{p,k}\bm{v}_{\bm{k}}\bm{v}_{\bm{p}}[1-F^2(x,k)][1-F^2(x',p)]\braket{\delta_G(x,k)\delta_G(x',p)}.
    \end{aligned}\label{eq:S3}
\end{equation}
It is apparent that in Eq.\eqref{eq:S3} the contribution at $k=p$ is canceled by the second part of Eq.\eqref{eq:S2}, so the total result is well regularized and we only need to consider the contributions with $k\neq p$ in Eq.\eqref{eq:S3}. Therefore, the total noise power can be calculated as 
\begin{equation}
    \begin{aligned}
        \mathcal{S}(T,V)&=\frac{e^2W^2}{2\Omega^2}\int d\br\sum_k \bm{v}_{\bk}^2[F^2(x,k)-1][G^R(x,k)-G^A(x,k)]^2\\
        &+\frac{e^2W^2}{2\Omega^2}\int d\br d\br' dt \sum_{p,k}\bm{v}_{\bm{k}}\bm{v}_{\bm{p}}[1-F^2(x,k)][1-F^2(x',p)]\braket{\delta_G(x,k)\delta_G(x',p)}
    \end{aligned}\label{eq:noisetotal}
\end{equation}
\end{widetext}
There are two contributions in this expression: the first one is the leading contribution and includes to Nagaev's treatment based on Boltzmann-Langivan formalism. To see this in a simple dirty Fermi gas system, one realizes that $G^R=[\omega-\epsilon+i\frac{1}{2\tau_{\text{im}}}]^{-1}$ and hence
\begin{equation}
    \int \frac{d\omega}{2\pi}(G^R-G^A)^2=-2\tau_{\text{im}}.
\end{equation}
If we approximate $F(x,k)$ by $F_s(x,k)$ which is identical to $1-2n_F(x,k)$ then the leading contribution reduces to\cite{NAGAEV1992103}
\begin{equation}
    \mathcal{S}=\frac{2e^2v_F^2\nu_0W}{L}\int_{-\frac{1}{2}}^{\frac{1}{2}}\frac{dx}{L}\int_{-\infty}^{+\infty}d\epsilon n_F(x,\epsilon)[1-n_F(x,\epsilon)]\tau_\text{im}.
\end{equation}
For the MFL scattering, or more generally, any scattering that has dependence on energy, to the leading accuracy one modifies the scattering time in the integrand properly and arrives at Eq.\eqref{eq:noisepower1}. Of course there are corrections, including the second line of Eq.\eqref{eq:noisetotal}, which are not considered here.

\bibliography{shotMFL}

\begin{thebibliography}{70}%
\makeatletter
\providecommand \@ifxundefined [1]{%
 \@ifx{#1\undefined}
}%
\providecommand \@ifnum [1]{%
 \ifnum #1\expandafter \@firstoftwo
 \else \expandafter \@secondoftwo
 \fi
}%
\providecommand \@ifx [1]{%
 \ifx #1\expandafter \@firstoftwo
 \else \expandafter \@secondoftwo
 \fi
}%
\providecommand \natexlab [1]{#1}%
\providecommand \enquote  [1]{``#1''}%
\providecommand \bibnamefont  [1]{#1}%
\providecommand \bibfnamefont [1]{#1}%
\providecommand \citenamefont [1]{#1}%
\providecommand \href@noop [0]{\@secondoftwo}%
\providecommand \href [0]{\begingroup \@sanitize@url \@href}%
\providecommand \@href[1]{\@@startlink{#1}\@@href}%
\providecommand \@@href[1]{\endgroup#1\@@endlink}%
\providecommand \@sanitize@url [0]{\catcode `\\12\catcode `\$12\catcode
  `\&12\catcode `\#12\catcode `\^12\catcode `\_12\catcode `\%12\relax}%
\providecommand \@@startlink[1]{}%
\providecommand \@@endlink[0]{}%
\providecommand \url  [0]{\begingroup\@sanitize@url \@url }%
\providecommand \@url [1]{\endgroup\@href {#1}{\urlprefix }}%
\providecommand \urlprefix  [0]{URL }%
\providecommand \Eprint [0]{\href }%
\providecommand \doibase [0]{https://doi.org/}%
\providecommand \selectlanguage [0]{\@gobble}%
\providecommand \bibinfo  [0]{\@secondoftwo}%
\providecommand \bibfield  [0]{\@secondoftwo}%
\providecommand \translation [1]{[#1]}%
\providecommand \BibitemOpen [0]{}%
\providecommand \bibitemStop [0]{}%
\providecommand \bibitemNoStop [0]{.\EOS\space}%
\providecommand \EOS [0]{\spacefactor3000\relax}%
\providecommand \BibitemShut  [1]{\csname bibitem#1\endcsname}%
\let\auto@bib@innerbib\@empty
\bibitem [{\citenamefont {Takagi}\ \emph {et~al.}(1992)\citenamefont {Takagi},
  \citenamefont {Batlogg}, \citenamefont {Kao}, \citenamefont {Kwo},
  \citenamefont {Cava}, \citenamefont {Krajewski},\ and\ \citenamefont
  {Peck}}]{PhysRevLett.69.2975}%
  \BibitemOpen
  \bibfield  {author} {\bibinfo {author} {\bibfnamefont {H.}~\bibnamefont
  {Takagi}}, \bibinfo {author} {\bibfnamefont {B.}~\bibnamefont {Batlogg}},
  \bibinfo {author} {\bibfnamefont {H.~L.}\ \bibnamefont {Kao}}, \bibinfo
  {author} {\bibfnamefont {J.}~\bibnamefont {Kwo}}, \bibinfo {author}
  {\bibfnamefont {R.~J.}\ \bibnamefont {Cava}}, \bibinfo {author}
  {\bibfnamefont {J.~J.}\ \bibnamefont {Krajewski}},\ and\ \bibinfo {author}
  {\bibfnamefont {W.~F.}\ \bibnamefont {Peck}},\ }\bibfield  {title} {\bibinfo
  {title} {Systematic evolution of temperature-dependent resistivity in
  ${\mathrm{la}}_{2\mathrm{\ensuremath{-}}\mathit{x}}$${\mathrm{sr}}_{\mathit{x}}$${\mathrm{cuo}}_{4}$},\
  }\href {https://doi.org/10.1103/PhysRevLett.69.2975} {\bibfield  {journal}
  {\bibinfo  {journal} {Phys. Rev. Lett.}\ }\textbf {\bibinfo {volume} {69}},\
  \bibinfo {pages} {2975} (\bibinfo {year} {1992})}\BibitemShut {NoStop}%
\bibitem [{\citenamefont {Gurvitch}\ and\ \citenamefont
  {Fiory}(1987)}]{PhysRevLett.59.1337}%
  \BibitemOpen
  \bibfield  {author} {\bibinfo {author} {\bibfnamefont {M.}~\bibnamefont
  {Gurvitch}}\ and\ \bibinfo {author} {\bibfnamefont {A.~T.}\ \bibnamefont
  {Fiory}},\ }\bibfield  {title} {\bibinfo {title} {Resistivity of
  ${\mathrm{la}}_{1.825}$${\mathrm{sr}}_{0.175}$${\mathrm{cuo}}_{4}$ and
  ${\mathrm{yba}}_{2}$${\mathrm{cu}}_{3}$${\mathrm{o}}_{7}$ to 1100 k: Absence
  of saturation and its implications},\ }\href
  {https://doi.org/10.1103/PhysRevLett.59.1337} {\bibfield  {journal} {\bibinfo
   {journal} {Phys. Rev. Lett.}\ }\textbf {\bibinfo {volume} {59}},\ \bibinfo
  {pages} {1337} (\bibinfo {year} {1987})}\BibitemShut {NoStop}%
\bibitem [{\citenamefont {Keimer}\ \emph {et~al.}(2015)\citenamefont {Keimer},
  \citenamefont {Kivelson}, \citenamefont {Norman}, \citenamefont {Uchida},\
  and\ \citenamefont {Zaanen}}]{Keimer2015}%
  \BibitemOpen
  \bibfield  {author} {\bibinfo {author} {\bibfnamefont {B.}~\bibnamefont
  {Keimer}}, \bibinfo {author} {\bibfnamefont {S.~A.}\ \bibnamefont
  {Kivelson}}, \bibinfo {author} {\bibfnamefont {M.~R.}\ \bibnamefont
  {Norman}}, \bibinfo {author} {\bibfnamefont {S.}~\bibnamefont {Uchida}},\
  and\ \bibinfo {author} {\bibfnamefont {J.}~\bibnamefont {Zaanen}},\
  }\bibfield  {title} {\bibinfo {title} {From quantum matter to
  high-temperature superconductivity in copper oxides},\ }\href
  {https://doi.org/10.1038/nature14165} {\bibfield  {journal} {\bibinfo
  {journal} {Nature}\ }\textbf {\bibinfo {volume} {518}},\ \bibinfo {pages}
  {179} (\bibinfo {year} {2015})}\BibitemShut {NoStop}%
\bibitem [{\citenamefont {Zaanen}(2019)}]{10.21468/SciPostPhys.6.5.061}%
  \BibitemOpen
  \bibfield  {author} {\bibinfo {author} {\bibfnamefont {J.}~\bibnamefont
  {Zaanen}},\ }\bibfield  {title} {\bibinfo {title} {{Planckian dissipation,
  minimal viscosity and the transport in cuprate strange metals}},\ }\href
  {https://doi.org/10.21468/SciPostPhys.6.5.061} {\bibfield  {journal}
  {\bibinfo  {journal} {SciPost Phys.}\ }\textbf {\bibinfo {volume} {6}},\
  \bibinfo {pages} {061} (\bibinfo {year} {2019})}\BibitemShut {NoStop}%
\bibitem [{\citenamefont {Varma}(2020)}]{RevModPhys.92.031001}%
  \BibitemOpen
  \bibfield  {author} {\bibinfo {author} {\bibfnamefont {C.~M.}\ \bibnamefont
  {Varma}},\ }\bibfield  {title} {\bibinfo {title} {Colloquium: Linear in
  temperature resistivity and associated mysteries including high temperature
  superconductivity},\ }\href {https://doi.org/10.1103/RevModPhys.92.031001}
  {\bibfield  {journal} {\bibinfo  {journal} {Rev. Mod. Phys.}\ }\textbf
  {\bibinfo {volume} {92}},\ \bibinfo {pages} {031001} (\bibinfo {year}
  {2020})}\BibitemShut {NoStop}%
\bibitem [{\citenamefont {Yuan}\ \emph {et~al.}(2022)\citenamefont {Yuan},
  \citenamefont {Chen}, \citenamefont {Jiang}, \citenamefont {Feng},
  \citenamefont {Lin}, \citenamefont {Yu}, \citenamefont {He}, \citenamefont
  {Zhang}, \citenamefont {Jiang}, \citenamefont {Zhang}, \citenamefont {Shi},
  \citenamefont {Zhang}, \citenamefont {Qin}, \citenamefont {Cheng},
  \citenamefont {Tamura}, \citenamefont {Yang}, \citenamefont {Xiang},
  \citenamefont {Hu}, \citenamefont {Takeuchi}, \citenamefont {Jin},\ and\
  \citenamefont {Zhao}}]{Yuan2022}%
  \BibitemOpen
  \bibfield  {author} {\bibinfo {author} {\bibfnamefont {J.}~\bibnamefont
  {Yuan}}, \bibinfo {author} {\bibfnamefont {Q.}~\bibnamefont {Chen}}, \bibinfo
  {author} {\bibfnamefont {K.}~\bibnamefont {Jiang}}, \bibinfo {author}
  {\bibfnamefont {Z.}~\bibnamefont {Feng}}, \bibinfo {author} {\bibfnamefont
  {Z.}~\bibnamefont {Lin}}, \bibinfo {author} {\bibfnamefont {H.}~\bibnamefont
  {Yu}}, \bibinfo {author} {\bibfnamefont {G.}~\bibnamefont {He}}, \bibinfo
  {author} {\bibfnamefont {J.}~\bibnamefont {Zhang}}, \bibinfo {author}
  {\bibfnamefont {X.}~\bibnamefont {Jiang}}, \bibinfo {author} {\bibfnamefont
  {X.}~\bibnamefont {Zhang}}, \bibinfo {author} {\bibfnamefont
  {Y.}~\bibnamefont {Shi}}, \bibinfo {author} {\bibfnamefont {Y.}~\bibnamefont
  {Zhang}}, \bibinfo {author} {\bibfnamefont {M.}~\bibnamefont {Qin}}, \bibinfo
  {author} {\bibfnamefont {Z.~G.}\ \bibnamefont {Cheng}}, \bibinfo {author}
  {\bibfnamefont {N.}~\bibnamefont {Tamura}}, \bibinfo {author} {\bibfnamefont
  {Y.-f.}\ \bibnamefont {Yang}}, \bibinfo {author} {\bibfnamefont
  {T.}~\bibnamefont {Xiang}}, \bibinfo {author} {\bibfnamefont
  {J.}~\bibnamefont {Hu}}, \bibinfo {author} {\bibfnamefont {I.}~\bibnamefont
  {Takeuchi}}, \bibinfo {author} {\bibfnamefont {K.}~\bibnamefont {Jin}},\ and\
  \bibinfo {author} {\bibfnamefont {Z.}~\bibnamefont {Zhao}},\ }\bibfield
  {title} {\bibinfo {title} {Scaling of the strange-metal scattering in
  unconventional superconductors},\ }\href
  {https://doi.org/10.1038/s41586-021-04305-5} {\bibfield  {journal} {\bibinfo
  {journal} {Nature}\ }\textbf {\bibinfo {volume} {602}},\ \bibinfo {pages}
  {431} (\bibinfo {year} {2022})}\BibitemShut {NoStop}%
\bibitem [{\citenamefont {von Löhneysen}(1996)}]{Hilbert996}%
  \BibitemOpen
  \bibfield  {author} {\bibinfo {author} {\bibfnamefont {H.}~\bibnamefont {von
  Löhneysen}},\ }\bibfield  {title} {\bibinfo {title} {Non-fermi-liquid
  behaviour in the heavy-fermion system},\ }\href
  {https://doi.org/10.1088/0953-8984/8/48/003} {\bibfield  {journal} {\bibinfo
  {journal} {Journal of Physics: Condensed Matter}\ }\textbf {\bibinfo {volume}
  {8}},\ \bibinfo {pages} {9689} (\bibinfo {year} {1996})}\BibitemShut
  {NoStop}%
\bibitem [{\citenamefont {L\"ohneysen}\ \emph {et~al.}(1994)\citenamefont
  {L\"ohneysen}, \citenamefont {Pietrus}, \citenamefont {Portisch},
  \citenamefont {Schlager}, \citenamefont {Schr\"oder}, \citenamefont {Sieck},\
  and\ \citenamefont {Trappmann}}]{PhysRevLett.72.3262}%
  \BibitemOpen
  \bibfield  {author} {\bibinfo {author} {\bibfnamefont {H.~v.}\ \bibnamefont
  {L\"ohneysen}}, \bibinfo {author} {\bibfnamefont {T.}~\bibnamefont
  {Pietrus}}, \bibinfo {author} {\bibfnamefont {G.}~\bibnamefont {Portisch}},
  \bibinfo {author} {\bibfnamefont {H.~G.}\ \bibnamefont {Schlager}}, \bibinfo
  {author} {\bibfnamefont {A.}~\bibnamefont {Schr\"oder}}, \bibinfo {author}
  {\bibfnamefont {M.}~\bibnamefont {Sieck}},\ and\ \bibinfo {author}
  {\bibfnamefont {T.}~\bibnamefont {Trappmann}},\ }\bibfield  {title} {\bibinfo
  {title} {Non-fermi-liquid behavior in a heavy-fermion alloy at a magnetic
  instability},\ }\href {https://doi.org/10.1103/PhysRevLett.72.3262}
  {\bibfield  {journal} {\bibinfo  {journal} {Phys. Rev. Lett.}\ }\textbf
  {\bibinfo {volume} {72}},\ \bibinfo {pages} {3262} (\bibinfo {year}
  {1994})}\BibitemShut {NoStop}%
\bibitem [{\citenamefont {Park}\ \emph {et~al.}(2006)\citenamefont {Park},
  \citenamefont {Ronning}, \citenamefont {Yuan}, \citenamefont {Salamon},
  \citenamefont {Movshovich}, \citenamefont {Sarrao},\ and\ \citenamefont
  {Thompson}}]{Park2006}%
  \BibitemOpen
  \bibfield  {author} {\bibinfo {author} {\bibfnamefont {T.}~\bibnamefont
  {Park}}, \bibinfo {author} {\bibfnamefont {F.}~\bibnamefont {Ronning}},
  \bibinfo {author} {\bibfnamefont {H.~Q.}\ \bibnamefont {Yuan}}, \bibinfo
  {author} {\bibfnamefont {M.~B.}\ \bibnamefont {Salamon}}, \bibinfo {author}
  {\bibfnamefont {R.}~\bibnamefont {Movshovich}}, \bibinfo {author}
  {\bibfnamefont {J.~L.}\ \bibnamefont {Sarrao}},\ and\ \bibinfo {author}
  {\bibfnamefont {J.~D.}\ \bibnamefont {Thompson}},\ }\bibfield  {title}
  {\bibinfo {title} {Hidden magnetism and quantum criticality in the heavy
  fermion superconductor cerhin5},\ }\href
  {https://doi.org/10.1038/nature04571} {\bibfield  {journal} {\bibinfo
  {journal} {Nature}\ }\textbf {\bibinfo {volume} {440}},\ \bibinfo {pages}
  {65} (\bibinfo {year} {2006})}\BibitemShut {NoStop}%
\bibitem [{\citenamefont {Trovarelli}\ \emph {et~al.}(2000)\citenamefont
  {Trovarelli}, \citenamefont {Geibel}, \citenamefont {Mederle}, \citenamefont
  {Langhammer}, \citenamefont {Grosche}, \citenamefont {Gegenwart},
  \citenamefont {Lang}, \citenamefont {Sparn},\ and\ \citenamefont
  {Steglich}}]{PhysRevLett.85.626}%
  \BibitemOpen
  \bibfield  {author} {\bibinfo {author} {\bibfnamefont {O.}~\bibnamefont
  {Trovarelli}}, \bibinfo {author} {\bibfnamefont {C.}~\bibnamefont {Geibel}},
  \bibinfo {author} {\bibfnamefont {S.}~\bibnamefont {Mederle}}, \bibinfo
  {author} {\bibfnamefont {C.}~\bibnamefont {Langhammer}}, \bibinfo {author}
  {\bibfnamefont {F.~M.}\ \bibnamefont {Grosche}}, \bibinfo {author}
  {\bibfnamefont {P.}~\bibnamefont {Gegenwart}}, \bibinfo {author}
  {\bibfnamefont {M.}~\bibnamefont {Lang}}, \bibinfo {author} {\bibfnamefont
  {G.}~\bibnamefont {Sparn}},\ and\ \bibinfo {author} {\bibfnamefont
  {F.}~\bibnamefont {Steglich}},\ }\bibfield  {title} {\bibinfo {title}
  {${\mathrm{ybrh}}_{2}{\mathrm{si}}_{2}$: Pronounced non-fermi-liquid effects
  above a low-lying magnetic phase transition},\ }\href
  {https://doi.org/10.1103/PhysRevLett.85.626} {\bibfield  {journal} {\bibinfo
  {journal} {Phys. Rev. Lett.}\ }\textbf {\bibinfo {volume} {85}},\ \bibinfo
  {pages} {626} (\bibinfo {year} {2000})}\BibitemShut {NoStop}%
\bibitem [{\citenamefont {Gegenwart}\ \emph {et~al.}(2002)\citenamefont
  {Gegenwart}, \citenamefont {Custers}, \citenamefont {Geibel}, \citenamefont
  {Neumaier}, \citenamefont {Tayama}, \citenamefont {Tenya}, \citenamefont
  {Trovarelli},\ and\ \citenamefont {Steglich}}]{PhysRevLett.89.056402}%
  \BibitemOpen
  \bibfield  {author} {\bibinfo {author} {\bibfnamefont {P.}~\bibnamefont
  {Gegenwart}}, \bibinfo {author} {\bibfnamefont {J.}~\bibnamefont {Custers}},
  \bibinfo {author} {\bibfnamefont {C.}~\bibnamefont {Geibel}}, \bibinfo
  {author} {\bibfnamefont {K.}~\bibnamefont {Neumaier}}, \bibinfo {author}
  {\bibfnamefont {T.}~\bibnamefont {Tayama}}, \bibinfo {author} {\bibfnamefont
  {K.}~\bibnamefont {Tenya}}, \bibinfo {author} {\bibfnamefont
  {O.}~\bibnamefont {Trovarelli}},\ and\ \bibinfo {author} {\bibfnamefont
  {F.}~\bibnamefont {Steglich}},\ }\bibfield  {title} {\bibinfo {title}
  {Magnetic-field induced quantum critical point in
  $\mathrm{Y}\mathrm{b}\mathrm{R}{\mathrm{h}}_{\mathrm{2}}\mathrm{S}{\mathrm{i}}_{\mathrm{2}}$},\
  }\href {https://doi.org/10.1103/PhysRevLett.89.056402} {\bibfield  {journal}
  {\bibinfo  {journal} {Phys. Rev. Lett.}\ }\textbf {\bibinfo {volume} {89}},\
  \bibinfo {pages} {056402} (\bibinfo {year} {2002})}\BibitemShut {NoStop}%
\bibitem [{\citenamefont {Kasahara}\ \emph {et~al.}(2010)\citenamefont
  {Kasahara}, \citenamefont {Shibauchi}, \citenamefont {Hashimoto},
  \citenamefont {Ikada}, \citenamefont {Tonegawa}, \citenamefont {Okazaki},
  \citenamefont {Shishido}, \citenamefont {Ikeda}, \citenamefont {Takeya},
  \citenamefont {Hirata}, \citenamefont {Terashima},\ and\ \citenamefont
  {Matsuda}}]{PhysRevB.81.184519}%
  \BibitemOpen
  \bibfield  {author} {\bibinfo {author} {\bibfnamefont {S.}~\bibnamefont
  {Kasahara}}, \bibinfo {author} {\bibfnamefont {T.}~\bibnamefont {Shibauchi}},
  \bibinfo {author} {\bibfnamefont {K.}~\bibnamefont {Hashimoto}}, \bibinfo
  {author} {\bibfnamefont {K.}~\bibnamefont {Ikada}}, \bibinfo {author}
  {\bibfnamefont {S.}~\bibnamefont {Tonegawa}}, \bibinfo {author}
  {\bibfnamefont {R.}~\bibnamefont {Okazaki}}, \bibinfo {author} {\bibfnamefont
  {H.}~\bibnamefont {Shishido}}, \bibinfo {author} {\bibfnamefont
  {H.}~\bibnamefont {Ikeda}}, \bibinfo {author} {\bibfnamefont
  {H.}~\bibnamefont {Takeya}}, \bibinfo {author} {\bibfnamefont
  {K.}~\bibnamefont {Hirata}}, \bibinfo {author} {\bibfnamefont
  {T.}~\bibnamefont {Terashima}},\ and\ \bibinfo {author} {\bibfnamefont
  {Y.}~\bibnamefont {Matsuda}},\ }\bibfield  {title} {\bibinfo {title}
  {Evolution from non-fermi- to fermi-liquid transport via isovalent doping in
  ${\text{bafe}}_{2}{({\text{As}}_{1\ensuremath{-}x}{\text{P}}_{x})}_{2}$
  superconductors},\ }\href {https://doi.org/10.1103/PhysRevB.81.184519}
  {\bibfield  {journal} {\bibinfo  {journal} {Phys. Rev. B}\ }\textbf {\bibinfo
  {volume} {81}},\ \bibinfo {pages} {184519} (\bibinfo {year}
  {2010})}\BibitemShut {NoStop}%
\bibitem [{\citenamefont {Hayes}\ \emph {et~al.}(2016)\citenamefont {Hayes},
  \citenamefont {McDonald}, \citenamefont {Breznay}, \citenamefont {Helm},
  \citenamefont {Moll}, \citenamefont {Wartenbe}, \citenamefont {Shekhter},\
  and\ \citenamefont {Analytis}}]{Hayes2016}%
  \BibitemOpen
  \bibfield  {author} {\bibinfo {author} {\bibfnamefont {I.~M.}\ \bibnamefont
  {Hayes}}, \bibinfo {author} {\bibfnamefont {R.~D.}\ \bibnamefont {McDonald}},
  \bibinfo {author} {\bibfnamefont {N.~P.}\ \bibnamefont {Breznay}}, \bibinfo
  {author} {\bibfnamefont {T.}~\bibnamefont {Helm}}, \bibinfo {author}
  {\bibfnamefont {P.~J.~W.}\ \bibnamefont {Moll}}, \bibinfo {author}
  {\bibfnamefont {M.}~\bibnamefont {Wartenbe}}, \bibinfo {author}
  {\bibfnamefont {A.}~\bibnamefont {Shekhter}},\ and\ \bibinfo {author}
  {\bibfnamefont {J.~G.}\ \bibnamefont {Analytis}},\ }\bibfield  {title}
  {\bibinfo {title} {Scaling between magnetic field and temperature in the
  high-temperature superconductor ${BaFe}_{2}$$({As}_{1-x}{P}_{x})2$},\ }\href
  {https://doi.org/10.1038/nphys3773} {\bibfield  {journal} {\bibinfo
  {journal} {Nature Physics}\ }\textbf {\bibinfo {volume} {12}},\ \bibinfo
  {pages} {916} (\bibinfo {year} {2016})}\BibitemShut {NoStop}%
\bibitem [{\citenamefont {Shibauchi}\ \emph {et~al.}(2014)\citenamefont
  {Shibauchi}, \citenamefont {Carrington},\ and\ \citenamefont
  {Matsuda}}]{annurev1}%
  \BibitemOpen
  \bibfield  {author} {\bibinfo {author} {\bibfnamefont {T.}~\bibnamefont
  {Shibauchi}}, \bibinfo {author} {\bibfnamefont {A.}~\bibnamefont
  {Carrington}},\ and\ \bibinfo {author} {\bibfnamefont {Y.}~\bibnamefont
  {Matsuda}},\ }\bibfield  {title} {\bibinfo {title} {A quantum critical point
  lying beneath the superconducting dome in iron pnictides},\ }\href
  {https://doi.org/https://doi.org/10.1146/annurev-conmatphys-031113-133921}
  {\bibfield  {journal} {\bibinfo  {journal} {Annual Review of Condensed Matter
  Physics}\ }\textbf {\bibinfo {volume} {5}},\ \bibinfo {pages} {113} (\bibinfo
  {year} {2014})}\BibitemShut {NoStop}%
\bibitem [{\citenamefont {Jiang}\ \emph {et~al.}(2023)\citenamefont {Jiang},
  \citenamefont {Qin}, \citenamefont {Wei}, \citenamefont {Xu}, \citenamefont
  {Ke}, \citenamefont {Zhu}, \citenamefont {Zhang}, \citenamefont {Zhao},
  \citenamefont {Liang}, \citenamefont {Wei}, \citenamefont {Lin},
  \citenamefont {Feng}, \citenamefont {Chen}, \citenamefont {Xiong},
  \citenamefont {Yuan}, \citenamefont {Zhu}, \citenamefont {Li}, \citenamefont
  {Xi}, \citenamefont {Wang}, \citenamefont {Yang}, \citenamefont {Wang},
  \citenamefont {Xiang}, \citenamefont {Hu}, \citenamefont {Jiang},
  \citenamefont {Chen}, \citenamefont {Jin},\ and\ \citenamefont
  {Zhao}}]{Jiang2023}%
  \BibitemOpen
  \bibfield  {author} {\bibinfo {author} {\bibfnamefont {X.}~\bibnamefont
  {Jiang}}, \bibinfo {author} {\bibfnamefont {M.}~\bibnamefont {Qin}}, \bibinfo
  {author} {\bibfnamefont {X.}~\bibnamefont {Wei}}, \bibinfo {author}
  {\bibfnamefont {L.}~\bibnamefont {Xu}}, \bibinfo {author} {\bibfnamefont
  {J.}~\bibnamefont {Ke}}, \bibinfo {author} {\bibfnamefont {H.}~\bibnamefont
  {Zhu}}, \bibinfo {author} {\bibfnamefont {R.}~\bibnamefont {Zhang}}, \bibinfo
  {author} {\bibfnamefont {Z.}~\bibnamefont {Zhao}}, \bibinfo {author}
  {\bibfnamefont {Q.}~\bibnamefont {Liang}}, \bibinfo {author} {\bibfnamefont
  {Z.}~\bibnamefont {Wei}}, \bibinfo {author} {\bibfnamefont {Z.}~\bibnamefont
  {Lin}}, \bibinfo {author} {\bibfnamefont {Z.}~\bibnamefont {Feng}}, \bibinfo
  {author} {\bibfnamefont {F.}~\bibnamefont {Chen}}, \bibinfo {author}
  {\bibfnamefont {P.}~\bibnamefont {Xiong}}, \bibinfo {author} {\bibfnamefont
  {J.}~\bibnamefont {Yuan}}, \bibinfo {author} {\bibfnamefont {B.}~\bibnamefont
  {Zhu}}, \bibinfo {author} {\bibfnamefont {Y.}~\bibnamefont {Li}}, \bibinfo
  {author} {\bibfnamefont {C.}~\bibnamefont {Xi}}, \bibinfo {author}
  {\bibfnamefont {Z.}~\bibnamefont {Wang}}, \bibinfo {author} {\bibfnamefont
  {M.}~\bibnamefont {Yang}}, \bibinfo {author} {\bibfnamefont {J.}~\bibnamefont
  {Wang}}, \bibinfo {author} {\bibfnamefont {T.}~\bibnamefont {Xiang}},
  \bibinfo {author} {\bibfnamefont {J.}~\bibnamefont {Hu}}, \bibinfo {author}
  {\bibfnamefont {K.}~\bibnamefont {Jiang}}, \bibinfo {author} {\bibfnamefont
  {Q.}~\bibnamefont {Chen}}, \bibinfo {author} {\bibfnamefont {K.}~\bibnamefont
  {Jin}},\ and\ \bibinfo {author} {\bibfnamefont {Z.}~\bibnamefont {Zhao}},\
  }\bibfield  {title} {\bibinfo {title} {Interplay between superconductivity
  and the strange-metal state in fese},\ }\href
  {https://doi.org/10.1038/s41567-022-01894-4} {\bibfield  {journal} {\bibinfo
  {journal} {Nature Physics}\ }\textbf {\bibinfo {volume} {19}},\ \bibinfo
  {pages} {365} (\bibinfo {year} {2023})}\BibitemShut {NoStop}%
\bibitem [{\citenamefont {Cao}\ \emph {et~al.}(2020)\citenamefont {Cao},
  \citenamefont {Chowdhury}, \citenamefont {Rodan-Legrain}, \citenamefont
  {Rubies-Bigorda}, \citenamefont {Watanabe}, \citenamefont {Taniguchi},
  \citenamefont {Senthil},\ and\ \citenamefont
  {Jarillo-Herrero}}]{PhysRevLett.124.076801}%
  \BibitemOpen
  \bibfield  {author} {\bibinfo {author} {\bibfnamefont {Y.}~\bibnamefont
  {Cao}}, \bibinfo {author} {\bibfnamefont {D.}~\bibnamefont {Chowdhury}},
  \bibinfo {author} {\bibfnamefont {D.}~\bibnamefont {Rodan-Legrain}}, \bibinfo
  {author} {\bibfnamefont {O.}~\bibnamefont {Rubies-Bigorda}}, \bibinfo
  {author} {\bibfnamefont {K.}~\bibnamefont {Watanabe}}, \bibinfo {author}
  {\bibfnamefont {T.}~\bibnamefont {Taniguchi}}, \bibinfo {author}
  {\bibfnamefont {T.}~\bibnamefont {Senthil}},\ and\ \bibinfo {author}
  {\bibfnamefont {P.}~\bibnamefont {Jarillo-Herrero}},\ }\bibfield  {title}
  {\bibinfo {title} {Strange metal in magic-angle graphene with near planckian
  dissipation},\ }\href {https://doi.org/10.1103/PhysRevLett.124.076801}
  {\bibfield  {journal} {\bibinfo  {journal} {Phys. Rev. Lett.}\ }\textbf
  {\bibinfo {volume} {124}},\ \bibinfo {pages} {076801} (\bibinfo {year}
  {2020})}\BibitemShut {NoStop}%
\bibitem [{\citenamefont {Chen}\ \emph {et~al.}(2023)\citenamefont {Chen},
  \citenamefont {Lowder}, \citenamefont {Bakali}, \citenamefont {Andrews},
  \citenamefont {Schrenk}, \citenamefont {Waas}, \citenamefont {Svagera},
  \citenamefont {Eguchi}, \citenamefont {Prochaska}, \citenamefont {Wang},
  \citenamefont {Setty}, \citenamefont {Sur}, \citenamefont {Si}, \citenamefont
  {Paschen},\ and\ \citenamefont {Natelson}}]{doi:10.1126/science.abq6100}%
  \BibitemOpen
  \bibfield  {author} {\bibinfo {author} {\bibfnamefont {L.}~\bibnamefont
  {Chen}}, \bibinfo {author} {\bibfnamefont {D.~T.}\ \bibnamefont {Lowder}},
  \bibinfo {author} {\bibfnamefont {E.}~\bibnamefont {Bakali}}, \bibinfo
  {author} {\bibfnamefont {A.~M.}\ \bibnamefont {Andrews}}, \bibinfo {author}
  {\bibfnamefont {W.}~\bibnamefont {Schrenk}}, \bibinfo {author} {\bibfnamefont
  {M.}~\bibnamefont {Waas}}, \bibinfo {author} {\bibfnamefont {R.}~\bibnamefont
  {Svagera}}, \bibinfo {author} {\bibfnamefont {G.}~\bibnamefont {Eguchi}},
  \bibinfo {author} {\bibfnamefont {L.}~\bibnamefont {Prochaska}}, \bibinfo
  {author} {\bibfnamefont {Y.}~\bibnamefont {Wang}}, \bibinfo {author}
  {\bibfnamefont {C.}~\bibnamefont {Setty}}, \bibinfo {author} {\bibfnamefont
  {S.}~\bibnamefont {Sur}}, \bibinfo {author} {\bibfnamefont {Q.}~\bibnamefont
  {Si}}, \bibinfo {author} {\bibfnamefont {S.}~\bibnamefont {Paschen}},\ and\
  \bibinfo {author} {\bibfnamefont {D.}~\bibnamefont {Natelson}},\ }\bibfield
  {title} {\bibinfo {title} {Shot noise in a strange metal},\ }\href
  {https://doi.org/10.1126/science.abq6100} {\bibfield  {journal} {\bibinfo
  {journal} {Science}\ }\textbf {\bibinfo {volume} {382}},\ \bibinfo {pages}
  {907} (\bibinfo {year} {2023})}\BibitemShut {NoStop}%
\bibitem [{\citenamefont {Blanter}\ and\ \citenamefont
  {Büttiker}(2000)}]{BLANTER20001}%
  \BibitemOpen
  \bibfield  {author} {\bibinfo {author} {\bibfnamefont {Y.}~\bibnamefont
  {Blanter}}\ and\ \bibinfo {author} {\bibfnamefont {M.}~\bibnamefont
  {Büttiker}},\ }\bibfield  {title} {\bibinfo {title} {Shot noise in
  mesoscopic conductors},\ }\href
  {https://doi.org/https://doi.org/10.1016/S0370-1573(99)00123-4} {\bibfield
  {journal} {\bibinfo  {journal} {Physics Reports}\ }\textbf {\bibinfo {volume}
  {336}},\ \bibinfo {pages} {1} (\bibinfo {year} {2000})}\BibitemShut {NoStop}%
\bibitem [{\citenamefont {de~Jong}\ and\ \citenamefont
  {Beenakker}(1997)}]{deJong1997}%
  \BibitemOpen
  \bibfield  {author} {\bibinfo {author} {\bibfnamefont {M.~J.~M.}\
  \bibnamefont {de~Jong}}\ and\ \bibinfo {author} {\bibfnamefont {C.~W.~J.}\
  \bibnamefont {Beenakker}},\ }\bibinfo {title} {Shot noise in mesoscopic
  systems},\ in\ \href {https://doi.org/10.1007/978-94-015-8839-3_6} {\emph
  {\bibinfo {booktitle} {Mesoscopic Electron Transport}}},\ \bibinfo {editor}
  {edited by\ \bibinfo {editor} {\bibfnamefont {L.~L.}\ \bibnamefont {Sohn}},
  \bibinfo {editor} {\bibfnamefont {L.~P.}\ \bibnamefont {Kouwenhoven}},\ and\
  \bibinfo {editor} {\bibfnamefont {G.}~\bibnamefont {Sch{\"o}n}}}\ (\bibinfo
  {publisher} {Springer Netherlands},\ \bibinfo {address} {Dordrecht},\
  \bibinfo {year} {1997})\ pp.\ \bibinfo {pages} {225--258}\BibitemShut
  {NoStop}%
\bibitem [{\citenamefont {Landauer}(1993)}]{PhysRevB.47.16427}%
  \BibitemOpen
  \bibfield  {author} {\bibinfo {author} {\bibfnamefont {R.}~\bibnamefont
  {Landauer}},\ }\bibfield  {title} {\bibinfo {title} {Solid-state shot
  noise},\ }\href {https://doi.org/10.1103/PhysRevB.47.16427} {\bibfield
  {journal} {\bibinfo  {journal} {Phys. Rev. B}\ }\textbf {\bibinfo {volume}
  {47}},\ \bibinfo {pages} {16427} (\bibinfo {year} {1993})}\BibitemShut
  {NoStop}%
\bibitem [{\citenamefont {Kobayashi}\ and\ \citenamefont
  {Hashisaka}(2021)}]{doi:10.7566/JPSJ.90.102001}%
  \BibitemOpen
  \bibfield  {author} {\bibinfo {author} {\bibfnamefont {K.}~\bibnamefont
  {Kobayashi}}\ and\ \bibinfo {author} {\bibfnamefont {M.}~\bibnamefont
  {Hashisaka}},\ }\bibfield  {title} {\bibinfo {title} {Shot noise in
  mesoscopic systems: From single particles to quantum liquids},\ }\href
  {https://doi.org/10.7566/JPSJ.90.102001} {\bibfield  {journal} {\bibinfo
  {journal} {Journal of the Physical Society of Japan}\ }\textbf {\bibinfo
  {volume} {90}},\ \bibinfo {pages} {102001} (\bibinfo {year}
  {2021})}\BibitemShut {NoStop}%
\bibitem [{\citenamefont {Landauer}\ and\ \citenamefont
  {Martin}(1991)}]{LANDAUER1991167}%
  \BibitemOpen
  \bibfield  {author} {\bibinfo {author} {\bibfnamefont {R.}~\bibnamefont
  {Landauer}}\ and\ \bibinfo {author} {\bibfnamefont {T.}~\bibnamefont
  {Martin}},\ }\bibfield  {title} {\bibinfo {title} {Equilibrium and shot noise
  in mesoscopic systems},\ }\href
  {https://doi.org/https://doi.org/10.1016/0921-4526(91)90710-V} {\bibfield
  {journal} {\bibinfo  {journal} {Physica B: Condensed Matter}\ }\textbf
  {\bibinfo {volume} {175}},\ \bibinfo {pages} {167} (\bibinfo {year}
  {1991})},\ \bibinfo {note} {analogies in Optics and
  Micro-Electronics}\BibitemShut {NoStop}%
\bibitem [{\citenamefont {de~Picciotto}\ \emph {et~al.}(1997)\citenamefont
  {de~Picciotto}, \citenamefont {Reznikov}, \citenamefont {Heiblum},
  \citenamefont {Umansky}, \citenamefont {Bunin},\ and\ \citenamefont
  {Mahalu}}]{de-Picciotto1997}%
  \BibitemOpen
  \bibfield  {author} {\bibinfo {author} {\bibfnamefont {R.}~\bibnamefont
  {de~Picciotto}}, \bibinfo {author} {\bibfnamefont {M.}~\bibnamefont
  {Reznikov}}, \bibinfo {author} {\bibfnamefont {M.}~\bibnamefont {Heiblum}},
  \bibinfo {author} {\bibfnamefont {V.}~\bibnamefont {Umansky}}, \bibinfo
  {author} {\bibfnamefont {G.}~\bibnamefont {Bunin}},\ and\ \bibinfo {author}
  {\bibfnamefont {D.}~\bibnamefont {Mahalu}},\ }\bibfield  {title} {\bibinfo
  {title} {Direct observation of a fractional charge},\ }\href
  {https://doi.org/10.1038/38241} {\bibfield  {journal} {\bibinfo  {journal}
  {Nature}\ }\textbf {\bibinfo {volume} {389}},\ \bibinfo {pages} {162}
  (\bibinfo {year} {1997})}\BibitemShut {NoStop}%
\bibitem [{\citenamefont {Saminadayar}\ \emph {et~al.}(1997)\citenamefont
  {Saminadayar}, \citenamefont {Glattli}, \citenamefont {Jin},\ and\
  \citenamefont {Etienne}}]{PhysRevLett.79.2526}%
  \BibitemOpen
  \bibfield  {author} {\bibinfo {author} {\bibfnamefont {L.}~\bibnamefont
  {Saminadayar}}, \bibinfo {author} {\bibfnamefont {D.~C.}\ \bibnamefont
  {Glattli}}, \bibinfo {author} {\bibfnamefont {Y.}~\bibnamefont {Jin}},\ and\
  \bibinfo {author} {\bibfnamefont {B.}~\bibnamefont {Etienne}},\ }\bibfield
  {title} {\bibinfo {title} {Observation of the $\mathit{e}\mathit{/}3$
  fractionally charged laughlin quasiparticle},\ }\href
  {https://doi.org/10.1103/PhysRevLett.79.2526} {\bibfield  {journal} {\bibinfo
   {journal} {Phys. Rev. Lett.}\ }\textbf {\bibinfo {volume} {79}},\ \bibinfo
  {pages} {2526} (\bibinfo {year} {1997})}\BibitemShut {NoStop}%
\bibitem [{\citenamefont {Nilsson}\ \emph {et~al.}(2008)\citenamefont
  {Nilsson}, \citenamefont {Akhmerov},\ and\ \citenamefont
  {Beenakker}}]{PhysRevLett.101.120403}%
  \BibitemOpen
  \bibfield  {author} {\bibinfo {author} {\bibfnamefont {J.}~\bibnamefont
  {Nilsson}}, \bibinfo {author} {\bibfnamefont {A.~R.}\ \bibnamefont
  {Akhmerov}},\ and\ \bibinfo {author} {\bibfnamefont {C.~W.~J.}\ \bibnamefont
  {Beenakker}},\ }\bibfield  {title} {\bibinfo {title} {Splitting of a cooper
  pair by a pair of majorana bound states},\ }\href
  {https://doi.org/10.1103/PhysRevLett.101.120403} {\bibfield  {journal}
  {\bibinfo  {journal} {Phys. Rev. Lett.}\ }\textbf {\bibinfo {volume} {101}},\
  \bibinfo {pages} {120403} (\bibinfo {year} {2008})}\BibitemShut {NoStop}%
\bibitem [{\citenamefont {Liu}\ \emph {et~al.}(2013)\citenamefont {Liu},
  \citenamefont {Zhang},\ and\ \citenamefont {Law}}]{PhysRevB.88.064509}%
  \BibitemOpen
  \bibfield  {author} {\bibinfo {author} {\bibfnamefont {J.}~\bibnamefont
  {Liu}}, \bibinfo {author} {\bibfnamefont {F.-C.}\ \bibnamefont {Zhang}},\
  and\ \bibinfo {author} {\bibfnamefont {K.~T.}\ \bibnamefont {Law}},\
  }\bibfield  {title} {\bibinfo {title} {Majorana fermion induced nonlocal
  current correlations in spin-orbit coupled superconducting wires},\ }\href
  {https://doi.org/10.1103/PhysRevB.88.064509} {\bibfield  {journal} {\bibinfo
  {journal} {Phys. Rev. B}\ }\textbf {\bibinfo {volume} {88}},\ \bibinfo
  {pages} {064509} (\bibinfo {year} {2013})}\BibitemShut {NoStop}%
\bibitem [{\citenamefont {Danneau}\ \emph {et~al.}(2008)\citenamefont
  {Danneau}, \citenamefont {Wu}, \citenamefont {Craciun}, \citenamefont
  {Russo}, \citenamefont {Tomi}, \citenamefont {Salmilehto}, \citenamefont
  {Morpurgo},\ and\ \citenamefont {Hakonen}}]{PhysRevLett.100.196802}%
  \BibitemOpen
  \bibfield  {author} {\bibinfo {author} {\bibfnamefont {R.}~\bibnamefont
  {Danneau}}, \bibinfo {author} {\bibfnamefont {F.}~\bibnamefont {Wu}},
  \bibinfo {author} {\bibfnamefont {M.~F.}\ \bibnamefont {Craciun}}, \bibinfo
  {author} {\bibfnamefont {S.}~\bibnamefont {Russo}}, \bibinfo {author}
  {\bibfnamefont {M.~Y.}\ \bibnamefont {Tomi}}, \bibinfo {author}
  {\bibfnamefont {J.}~\bibnamefont {Salmilehto}}, \bibinfo {author}
  {\bibfnamefont {A.~F.}\ \bibnamefont {Morpurgo}},\ and\ \bibinfo {author}
  {\bibfnamefont {P.~J.}\ \bibnamefont {Hakonen}},\ }\bibfield  {title}
  {\bibinfo {title} {Shot noise in ballistic graphene},\ }\href
  {https://doi.org/10.1103/PhysRevLett.100.196802} {\bibfield  {journal}
  {\bibinfo  {journal} {Phys. Rev. Lett.}\ }\textbf {\bibinfo {volume} {100}},\
  \bibinfo {pages} {196802} (\bibinfo {year} {2008})}\BibitemShut {NoStop}%
\bibitem [{\citenamefont {Birk}\ \emph {et~al.}(1995)\citenamefont {Birk},
  \citenamefont {de~Jong},\ and\ \citenamefont
  {Sch\"onenberger}}]{PhysRevLett.75.1610}%
  \BibitemOpen
  \bibfield  {author} {\bibinfo {author} {\bibfnamefont {H.}~\bibnamefont
  {Birk}}, \bibinfo {author} {\bibfnamefont {M.~J.~M.}\ \bibnamefont
  {de~Jong}},\ and\ \bibinfo {author} {\bibfnamefont {C.}~\bibnamefont
  {Sch\"onenberger}},\ }\bibfield  {title} {\bibinfo {title} {Shot-noise
  suppression in the single-electron tunneling regime},\ }\href
  {https://doi.org/10.1103/PhysRevLett.75.1610} {\bibfield  {journal} {\bibinfo
   {journal} {Phys. Rev. Lett.}\ }\textbf {\bibinfo {volume} {75}},\ \bibinfo
  {pages} {1610} (\bibinfo {year} {1995})}\BibitemShut {NoStop}%
\bibitem [{\citenamefont {Steinbach}\ \emph {et~al.}(1996)\citenamefont
  {Steinbach}, \citenamefont {Martinis},\ and\ \citenamefont
  {Devoret}}]{PhysRevLett.76.3806}%
  \BibitemOpen
  \bibfield  {author} {\bibinfo {author} {\bibfnamefont {A.~H.}\ \bibnamefont
  {Steinbach}}, \bibinfo {author} {\bibfnamefont {J.~M.}\ \bibnamefont
  {Martinis}},\ and\ \bibinfo {author} {\bibfnamefont {M.~H.}\ \bibnamefont
  {Devoret}},\ }\bibfield  {title} {\bibinfo {title} {Observation of
  hot-electron shot noise in a metallic resistor},\ }\href
  {https://doi.org/10.1103/PhysRevLett.76.3806} {\bibfield  {journal} {\bibinfo
   {journal} {Phys. Rev. Lett.}\ }\textbf {\bibinfo {volume} {76}},\ \bibinfo
  {pages} {3806} (\bibinfo {year} {1996})}\BibitemShut {NoStop}%
\bibitem [{Note1()}]{Note1}%
  \BibitemOpen
  \bibinfo {note} {Exceptions include cases where spin-flip second order
  tunneling dominates\cite {PhysRevLett.95.146806}}\BibitemShut {NoStop}%
\bibitem [{\citenamefont {Nagaev}(1992)}]{NAGAEV1992103}%
  \BibitemOpen
  \bibfield  {author} {\bibinfo {author} {\bibfnamefont {K.}~\bibnamefont
  {Nagaev}},\ }\bibfield  {title} {\bibinfo {title} {On the shot noise in dirty
  metal contacts},\ }\href
  {https://doi.org/https://doi.org/10.1016/0375-9601(92)90814-3} {\bibfield
  {journal} {\bibinfo  {journal} {Physics Letters A}\ }\textbf {\bibinfo
  {volume} {169}},\ \bibinfo {pages} {103} (\bibinfo {year}
  {1992})}\BibitemShut {NoStop}%
\bibitem [{\citenamefont {Nagaev}(1995)}]{PhysRevB.52.4740}%
  \BibitemOpen
  \bibfield  {author} {\bibinfo {author} {\bibfnamefont {K.~E.}\ \bibnamefont
  {Nagaev}},\ }\bibfield  {title} {\bibinfo {title} {Influence of
  electron-electron scattering on shot noise in diffusive contacts},\ }\href
  {https://doi.org/10.1103/PhysRevB.52.4740} {\bibfield  {journal} {\bibinfo
  {journal} {Phys. Rev. B}\ }\textbf {\bibinfo {volume} {52}},\ \bibinfo
  {pages} {4740} (\bibinfo {year} {1995})}\BibitemShut {NoStop}%
\bibitem [{\citenamefont {Kozub}\ and\ \citenamefont
  {Rudin}(1995)}]{PhysRevB.52.7853}%
  \BibitemOpen
  \bibfield  {author} {\bibinfo {author} {\bibfnamefont {V.~I.}\ \bibnamefont
  {Kozub}}\ and\ \bibinfo {author} {\bibfnamefont {A.~M.}\ \bibnamefont
  {Rudin}},\ }\bibfield  {title} {\bibinfo {title} {Shot noise in mesoscopic
  diffusive conductors in the limit of strong electron-electron scattering},\
  }\href {https://doi.org/10.1103/PhysRevB.52.7853} {\bibfield  {journal}
  {\bibinfo  {journal} {Phys. Rev. B}\ }\textbf {\bibinfo {volume} {52}},\
  \bibinfo {pages} {7853} (\bibinfo {year} {1995})}\BibitemShut {NoStop}%
\bibitem [{\citenamefont {Nikolaenko}\ \emph {et~al.}(2023)\citenamefont
  {Nikolaenko}, \citenamefont {Sachdev},\ and\ \citenamefont
  {Patel}}]{PhysRevResearch.5.043143}%
  \BibitemOpen
  \bibfield  {author} {\bibinfo {author} {\bibfnamefont {A.}~\bibnamefont
  {Nikolaenko}}, \bibinfo {author} {\bibfnamefont {S.}~\bibnamefont
  {Sachdev}},\ and\ \bibinfo {author} {\bibfnamefont {A.~A.}\ \bibnamefont
  {Patel}},\ }\bibfield  {title} {\bibinfo {title} {Theory of shot noise in
  strange metals},\ }\href {https://doi.org/10.1103/PhysRevResearch.5.043143}
  {\bibfield  {journal} {\bibinfo  {journal} {Phys. Rev. Res.}\ }\textbf
  {\bibinfo {volume} {5}},\ \bibinfo {pages} {043143} (\bibinfo {year}
  {2023})}\BibitemShut {NoStop}%
\bibitem [{\citenamefont {Wu}\ and\ \citenamefont
  {Foster}(2023)}]{wu2023suppression}%
  \BibitemOpen
  \bibfield  {author} {\bibinfo {author} {\bibfnamefont {T.~C.}\ \bibnamefont
  {Wu}}\ and\ \bibinfo {author} {\bibfnamefont {M.~S.}\ \bibnamefont
  {Foster}},\ }\href@noop {} {\bibinfo {title} {Suppression of shot noise in a
  dirty marginal fermi liquid}} (\bibinfo {year} {2023}),\ \Eprint
  {https://arxiv.org/abs/2312.03071} {arXiv:2312.03071 [cond-mat.str-el]}
  \BibitemShut {NoStop}%
\bibitem [{\citenamefont {Wang}\ \emph {et~al.}(2024)\citenamefont {Wang},
  \citenamefont {Sur}, \citenamefont {Setty}, \citenamefont {Natelson},\ and\
  \citenamefont {Si}}]{wang2024shotnoisecoupledelectronboson}%
  \BibitemOpen
  \bibfield  {author} {\bibinfo {author} {\bibfnamefont {Y.}~\bibnamefont
  {Wang}}, \bibinfo {author} {\bibfnamefont {S.}~\bibnamefont {Sur}}, \bibinfo
  {author} {\bibfnamefont {C.}~\bibnamefont {Setty}}, \bibinfo {author}
  {\bibfnamefont {D.}~\bibnamefont {Natelson}},\ and\ \bibinfo {author}
  {\bibfnamefont {Q.}~\bibnamefont {Si}},\ }\href
  {https://arxiv.org/abs/2404.14515} {\bibinfo {title} {Shot noise in coupled
  electron-boson systems}} (\bibinfo {year} {2024}),\ \Eprint
  {https://arxiv.org/abs/2404.14515} {arXiv:2404.14515 [cond-mat.str-el]}
  \BibitemShut {NoStop}%
\bibitem [{\citenamefont {Raghu}\ and\ \citenamefont
  {Varma}(2024)}]{raghu2024shotnoisenearquantumcriticality}%
  \BibitemOpen
  \bibfield  {author} {\bibinfo {author} {\bibfnamefont {S.}~\bibnamefont
  {Raghu}}\ and\ \bibinfo {author} {\bibfnamefont {C.~M.}\ \bibnamefont
  {Varma}},\ }\href {https://arxiv.org/abs/2409.10798} {\bibinfo {title} {Shot
  noise near quantum-criticality}} (\bibinfo {year} {2024}),\ \Eprint
  {https://arxiv.org/abs/2409.10798} {arXiv:2409.10798 [cond-mat.str-el]}
  \BibitemShut {NoStop}%
\bibitem [{\citenamefont {Varma}\ \emph {et~al.}(1989)\citenamefont {Varma},
  \citenamefont {Littlewood}, \citenamefont {Schmitt-Rink}, \citenamefont
  {Abrahams},\ and\ \citenamefont {Ruckenstein}}]{PhysRevLett.63.1996}%
  \BibitemOpen
  \bibfield  {author} {\bibinfo {author} {\bibfnamefont {C.~M.}\ \bibnamefont
  {Varma}}, \bibinfo {author} {\bibfnamefont {P.~B.}\ \bibnamefont
  {Littlewood}}, \bibinfo {author} {\bibfnamefont {S.}~\bibnamefont
  {Schmitt-Rink}}, \bibinfo {author} {\bibfnamefont {E.}~\bibnamefont
  {Abrahams}},\ and\ \bibinfo {author} {\bibfnamefont {A.~E.}\ \bibnamefont
  {Ruckenstein}},\ }\bibfield  {title} {\bibinfo {title} {Phenomenology of the
  normal state of cu-o high-temperature superconductors},\ }\href
  {https://doi.org/10.1103/PhysRevLett.63.1996} {\bibfield  {journal} {\bibinfo
   {journal} {Phys. Rev. Lett.}\ }\textbf {\bibinfo {volume} {63}},\ \bibinfo
  {pages} {1996} (\bibinfo {year} {1989})}\BibitemShut {NoStop}%
\bibitem [{\citenamefont {Pelzer}(1991)}]{PhysRevB.44.293}%
  \BibitemOpen
  \bibfield  {author} {\bibinfo {author} {\bibfnamefont {F.}~\bibnamefont
  {Pelzer}},\ }\bibfield  {title} {\bibinfo {title} {Amplitude of the de
  haas--van alphen oscillations for a marginal fermi liquid},\ }\href
  {https://doi.org/10.1103/PhysRevB.44.293} {\bibfield  {journal} {\bibinfo
  {journal} {Phys. Rev. B}\ }\textbf {\bibinfo {volume} {44}},\ \bibinfo
  {pages} {293} (\bibinfo {year} {1991})}\BibitemShut {NoStop}%
\bibitem [{\citenamefont {Littlewood}\ and\ \citenamefont
  {Varma}(1992)}]{PhysRevB.46.405}%
  \BibitemOpen
  \bibfield  {author} {\bibinfo {author} {\bibfnamefont {P.~B.}\ \bibnamefont
  {Littlewood}}\ and\ \bibinfo {author} {\bibfnamefont {C.~M.}\ \bibnamefont
  {Varma}},\ }\bibfield  {title} {\bibinfo {title} {Phenomenology of the
  superconductive state of a marginal fermi liquid},\ }\href
  {https://doi.org/10.1103/PhysRevB.46.405} {\bibfield  {journal} {\bibinfo
  {journal} {Phys. Rev. B}\ }\textbf {\bibinfo {volume} {46}},\ \bibinfo
  {pages} {405} (\bibinfo {year} {1992})}\BibitemShut {NoStop}%
\bibitem [{\citenamefont {Littlewood}\ and\ \citenamefont
  {Varma}(1991)}]{10.1063/1.348195}%
  \BibitemOpen
  \bibfield  {author} {\bibinfo {author} {\bibfnamefont {P.~B.}\ \bibnamefont
  {Littlewood}}\ and\ \bibinfo {author} {\bibfnamefont {C.~M.}\ \bibnamefont
  {Varma}},\ }\bibfield  {title} {\bibinfo {title} {{Phenomenology of the
  normal and superconducting states of a marginal Fermi liquid (invited)}},\
  }\href {https://doi.org/10.1063/1.348195} {\bibfield  {journal} {\bibinfo
  {journal} {Journal of Applied Physics}\ }\textbf {\bibinfo {volume} {69}},\
  \bibinfo {pages} {4979} (\bibinfo {year} {1991})}\BibitemShut {NoStop}%
\bibitem [{\citenamefont {Kakehashi}\ and\ \citenamefont
  {Fulde}(2005)}]{PhysRevLett.94.156401}%
  \BibitemOpen
  \bibfield  {author} {\bibinfo {author} {\bibfnamefont {Y.}~\bibnamefont
  {Kakehashi}}\ and\ \bibinfo {author} {\bibfnamefont {P.}~\bibnamefont
  {Fulde}},\ }\bibfield  {title} {\bibinfo {title} {Marginal fermi liquid
  theory in the hubbard model},\ }\href
  {https://doi.org/10.1103/PhysRevLett.94.156401} {\bibfield  {journal}
  {\bibinfo  {journal} {Phys. Rev. Lett.}\ }\textbf {\bibinfo {volume} {94}},\
  \bibinfo {pages} {156401} (\bibinfo {year} {2005})}\BibitemShut {NoStop}%
\bibitem [{\citenamefont {Gonz\'alez}\ \emph {et~al.}(1999)\citenamefont
  {Gonz\'alez}, \citenamefont {Guinea},\ and\ \citenamefont
  {Vozmediano}}]{PhysRevB.59.R2474}%
  \BibitemOpen
  \bibfield  {author} {\bibinfo {author} {\bibfnamefont {J.}~\bibnamefont
  {Gonz\'alez}}, \bibinfo {author} {\bibfnamefont {F.}~\bibnamefont {Guinea}},\
  and\ \bibinfo {author} {\bibfnamefont {M.~A.~H.}\ \bibnamefont
  {Vozmediano}},\ }\bibfield  {title} {\bibinfo {title} {Marginal-fermi-liquid
  behavior from two-dimensional coulomb interaction},\ }\href
  {https://doi.org/10.1103/PhysRevB.59.R2474} {\bibfield  {journal} {\bibinfo
  {journal} {Phys. Rev. B}\ }\textbf {\bibinfo {volume} {59}},\ \bibinfo
  {pages} {R2474} (\bibinfo {year} {1999})}\BibitemShut {NoStop}%
\bibitem [{\citenamefont {Gonz\'alez}\ and\ \citenamefont
  {Stauber}(2020)}]{PhysRevLett.124.186801}%
  \BibitemOpen
  \bibfield  {author} {\bibinfo {author} {\bibfnamefont {J.}~\bibnamefont
  {Gonz\'alez}}\ and\ \bibinfo {author} {\bibfnamefont {T.}~\bibnamefont
  {Stauber}},\ }\bibfield  {title} {\bibinfo {title} {Marginal fermi liquid in
  twisted bilayer graphene},\ }\href
  {https://doi.org/10.1103/PhysRevLett.124.186801} {\bibfield  {journal}
  {\bibinfo  {journal} {Phys. Rev. Lett.}\ }\textbf {\bibinfo {volume} {124}},\
  \bibinfo {pages} {186801} (\bibinfo {year} {2020})}\BibitemShut {NoStop}%
\bibitem [{\citenamefont {Gonz\'alez}(2014)}]{PhysRevB.90.121107}%
  \BibitemOpen
  \bibfield  {author} {\bibinfo {author} {\bibfnamefont {J.}~\bibnamefont
  {Gonz\'alez}},\ }\bibfield  {title} {\bibinfo {title} {Marginal fermi liquid
  versus excitonic instability in three-dimensional dirac semimetals},\ }\href
  {https://doi.org/10.1103/PhysRevB.90.121107} {\bibfield  {journal} {\bibinfo
  {journal} {Phys. Rev. B}\ }\textbf {\bibinfo {volume} {90}},\ \bibinfo
  {pages} {121107} (\bibinfo {year} {2014})}\BibitemShut {NoStop}%
\bibitem [{\citenamefont {Nosov}\ \emph {et~al.}(2024)\citenamefont {Nosov},
  \citenamefont {Wu},\ and\ \citenamefont {Raghu}}]{PhysRevB.109.075107}%
  \BibitemOpen
  \bibfield  {author} {\bibinfo {author} {\bibfnamefont {P.~A.}\ \bibnamefont
  {Nosov}}, \bibinfo {author} {\bibfnamefont {Y.-M.}\ \bibnamefont {Wu}},\ and\
  \bibinfo {author} {\bibfnamefont {S.}~\bibnamefont {Raghu}},\ }\bibfield
  {title} {\bibinfo {title} {Entropy and de haas--van alphen oscillations of a
  three-dimensional marginal fermi liquid},\ }\href
  {https://doi.org/10.1103/PhysRevB.109.075107} {\bibfield  {journal} {\bibinfo
   {journal} {Phys. Rev. B}\ }\textbf {\bibinfo {volume} {109}},\ \bibinfo
  {pages} {075107} (\bibinfo {year} {2024})}\BibitemShut {NoStop}%
\bibitem [{\citenamefont {Zhang}\ and\ \citenamefont {Chen}(2023)}]{Zhang2023}%
  \BibitemOpen
  \bibfield  {author} {\bibinfo {author} {\bibfnamefont {X.-T.}\ \bibnamefont
  {Zhang}}\ and\ \bibinfo {author} {\bibfnamefont {G.}~\bibnamefont {Chen}},\
  }\bibfield  {title} {\bibinfo {title} {Infinite critical boson non-fermi
  liquid},\ }\href {https://doi.org/10.1038/s41535-023-00543-0} {\bibfield
  {journal} {\bibinfo  {journal} {npj Quantum Materials}\ }\textbf {\bibinfo
  {volume} {8}},\ \bibinfo {pages} {10} (\bibinfo {year} {2023})}\BibitemShut
  {NoStop}%
\bibitem [{\citenamefont {Pan}\ and\ \citenamefont
  {Zhang}(2024)}]{PAN2024116451}%
  \BibitemOpen
  \bibfield  {author} {\bibinfo {author} {\bibfnamefont {Z.}~\bibnamefont
  {Pan}}\ and\ \bibinfo {author} {\bibfnamefont {X.-T.}\ \bibnamefont
  {Zhang}},\ }\bibfield  {title} {\bibinfo {title} {Infinite critical boson
  induced non-fermi liquid in d=3-epsilon dimensions},\ }\href
  {https://doi.org/https://doi.org/10.1016/j.nuclphysb.2024.116451} {\bibfield
  {journal} {\bibinfo  {journal} {Nuclear Physics B}\ }\textbf {\bibinfo
  {volume} {999}},\ \bibinfo {pages} {116451} (\bibinfo {year}
  {2024})}\BibitemShut {NoStop}%
\bibitem [{\citenamefont {Bashan}\ \emph {et~al.}(2023)\citenamefont {Bashan},
  \citenamefont {Tulipman}, \citenamefont {Schmalian},\ and\ \citenamefont
  {Berg}}]{bashan2023tunable}%
  \BibitemOpen
  \bibfield  {author} {\bibinfo {author} {\bibfnamefont {N.}~\bibnamefont
  {Bashan}}, \bibinfo {author} {\bibfnamefont {E.}~\bibnamefont {Tulipman}},
  \bibinfo {author} {\bibfnamefont {J.}~\bibnamefont {Schmalian}},\ and\
  \bibinfo {author} {\bibfnamefont {E.}~\bibnamefont {Berg}},\ }\href@noop {}
  {\bibinfo {title} {Tunable non-fermi liquid phase from coupling to two-level
  systems}} (\bibinfo {year} {2023}),\ \Eprint
  {https://arxiv.org/abs/2310.07768} {arXiv:2310.07768 [cond-mat.str-el]}
  \BibitemShut {NoStop}%
\bibitem [{\citenamefont {Tulipman}\ \emph {et~al.}(2024)\citenamefont
  {Tulipman}, \citenamefont {Bashan}, \citenamefont {Schmalian},\ and\
  \citenamefont {Berg}}]{tulipman2024solvable}%
  \BibitemOpen
  \bibfield  {author} {\bibinfo {author} {\bibfnamefont {E.}~\bibnamefont
  {Tulipman}}, \bibinfo {author} {\bibfnamefont {N.}~\bibnamefont {Bashan}},
  \bibinfo {author} {\bibfnamefont {J.}~\bibnamefont {Schmalian}},\ and\
  \bibinfo {author} {\bibfnamefont {E.}~\bibnamefont {Berg}},\ }\href@noop {}
  {\bibinfo {title} {Solvable models of two-level systems coupled to itinerant
  electrons: Robust non-fermi liquid and quantum critical pairing}} (\bibinfo
  {year} {2024}),\ \Eprint {https://arxiv.org/abs/2404.06532} {arXiv:2404.06532
  [cond-mat.str-el]} \BibitemShut {NoStop}%
\bibitem [{\citenamefont {{S. Charfi-Kaddour}}\ \emph
  {et~al.}(1992)\citenamefont {{S. Charfi-Kaddour}}, \citenamefont {{R. J.
  Tarento}},\ and\ \citenamefont {{M. Héritier}}}]{refId0}%
  \BibitemOpen
  \bibfield  {author} {\bibinfo {author} {\bibnamefont {{S. Charfi-Kaddour}}},
  \bibinfo {author} {\bibnamefont {{R. J. Tarento}}},\ and\ \bibinfo {author}
  {\bibnamefont {{M. Héritier}}},\ }\bibfield  {title} {\bibinfo {title} {Spin
  fluctuation effects on a quasi-2d itinerant electron system : a microscopic
  model for the marginal fermi liquid},\ }\href
  {https://doi.org/10.1051/jp1:1992251} {\bibfield  {journal} {\bibinfo
  {journal} {J. Phys. I France}\ }\textbf {\bibinfo {volume} {2}},\ \bibinfo
  {pages} {1853} (\bibinfo {year} {1992})}\BibitemShut {NoStop}%
\bibitem [{\citenamefont {Efetov}(2015)}]{PhysRevB.91.045110}%
  \BibitemOpen
  \bibfield  {author} {\bibinfo {author} {\bibfnamefont {K.~B.}\ \bibnamefont
  {Efetov}},\ }\bibfield  {title} {\bibinfo {title} {Quantum criticality in two
  dimensions and marginal fermi liquid},\ }\href
  {https://doi.org/10.1103/PhysRevB.91.045110} {\bibfield  {journal} {\bibinfo
  {journal} {Phys. Rev. B}\ }\textbf {\bibinfo {volume} {91}},\ \bibinfo
  {pages} {045110} (\bibinfo {year} {2015})}\BibitemShut {NoStop}%
\bibitem [{\citenamefont {Chowdhury}\ \emph {et~al.}(2018)\citenamefont
  {Chowdhury}, \citenamefont {Werman}, \citenamefont {Berg},\ and\
  \citenamefont {Senthil}}]{PhysRevX.8.031024}%
  \BibitemOpen
  \bibfield  {author} {\bibinfo {author} {\bibfnamefont {D.}~\bibnamefont
  {Chowdhury}}, \bibinfo {author} {\bibfnamefont {Y.}~\bibnamefont {Werman}},
  \bibinfo {author} {\bibfnamefont {E.}~\bibnamefont {Berg}},\ and\ \bibinfo
  {author} {\bibfnamefont {T.}~\bibnamefont {Senthil}},\ }\bibfield  {title}
  {\bibinfo {title} {Translationally invariant non-fermi-liquid metals with
  critical fermi surfaces: Solvable models},\ }\href
  {https://doi.org/10.1103/PhysRevX.8.031024} {\bibfield  {journal} {\bibinfo
  {journal} {Phys. Rev. X}\ }\textbf {\bibinfo {volume} {8}},\ \bibinfo {pages}
  {031024} (\bibinfo {year} {2018})}\BibitemShut {NoStop}%
\bibitem [{\citenamefont {Chowdhury}\ \emph {et~al.}(2022)\citenamefont
  {Chowdhury}, \citenamefont {Georges}, \citenamefont {Parcollet},\ and\
  \citenamefont {Sachdev}}]{RevModPhys.94.035004}%
  \BibitemOpen
  \bibfield  {author} {\bibinfo {author} {\bibfnamefont {D.}~\bibnamefont
  {Chowdhury}}, \bibinfo {author} {\bibfnamefont {A.}~\bibnamefont {Georges}},
  \bibinfo {author} {\bibfnamefont {O.}~\bibnamefont {Parcollet}},\ and\
  \bibinfo {author} {\bibfnamefont {S.}~\bibnamefont {Sachdev}},\ }\bibfield
  {title} {\bibinfo {title} {Sachdev-ye-kitaev models and beyond: Window into
  non-fermi liquids},\ }\href {https://doi.org/10.1103/RevModPhys.94.035004}
  {\bibfield  {journal} {\bibinfo  {journal} {Rev. Mod. Phys.}\ }\textbf
  {\bibinfo {volume} {94}},\ \bibinfo {pages} {035004} (\bibinfo {year}
  {2022})}\BibitemShut {NoStop}%
\bibitem [{\citenamefont {Esterlis}\ \emph {et~al.}(2021)\citenamefont
  {Esterlis}, \citenamefont {Guo}, \citenamefont {Patel},\ and\ \citenamefont
  {Sachdev}}]{PhysRevB.103.235129}%
  \BibitemOpen
  \bibfield  {author} {\bibinfo {author} {\bibfnamefont {I.}~\bibnamefont
  {Esterlis}}, \bibinfo {author} {\bibfnamefont {H.}~\bibnamefont {Guo}},
  \bibinfo {author} {\bibfnamefont {A.~A.}\ \bibnamefont {Patel}},\ and\
  \bibinfo {author} {\bibfnamefont {S.}~\bibnamefont {Sachdev}},\ }\bibfield
  {title} {\bibinfo {title} {Large-$n$ theory of critical fermi surfaces},\
  }\href {https://doi.org/10.1103/PhysRevB.103.235129} {\bibfield  {journal}
  {\bibinfo  {journal} {Phys. Rev. B}\ }\textbf {\bibinfo {volume} {103}},\
  \bibinfo {pages} {235129} (\bibinfo {year} {2021})}\BibitemShut {NoStop}%
\bibitem [{\citenamefont {Guo}\ \emph {et~al.}(2022)\citenamefont {Guo},
  \citenamefont {Patel}, \citenamefont {Esterlis},\ and\ \citenamefont
  {Sachdev}}]{PhysRevB.106.115151}%
  \BibitemOpen
  \bibfield  {author} {\bibinfo {author} {\bibfnamefont {H.}~\bibnamefont
  {Guo}}, \bibinfo {author} {\bibfnamefont {A.~A.}\ \bibnamefont {Patel}},
  \bibinfo {author} {\bibfnamefont {I.}~\bibnamefont {Esterlis}},\ and\
  \bibinfo {author} {\bibfnamefont {S.}~\bibnamefont {Sachdev}},\ }\bibfield
  {title} {\bibinfo {title} {Large-$n$ theory of critical fermi surfaces. ii.
  conductivity},\ }\href {https://doi.org/10.1103/PhysRevB.106.115151}
  {\bibfield  {journal} {\bibinfo  {journal} {Phys. Rev. B}\ }\textbf {\bibinfo
  {volume} {106}},\ \bibinfo {pages} {115151} (\bibinfo {year}
  {2022})}\BibitemShut {NoStop}%
\bibitem [{\citenamefont {Gantsevich}\ \emph {et~al.}(1979)\citenamefont
  {Gantsevich}, \citenamefont {Gurevich},\ and\ \citenamefont
  {Katilius}}]{Gantsevich1979}%
  \BibitemOpen
  \bibfield  {author} {\bibinfo {author} {\bibfnamefont {S.~V.}\ \bibnamefont
  {Gantsevich}}, \bibinfo {author} {\bibfnamefont {V.~L.}\ \bibnamefont
  {Gurevich}},\ and\ \bibinfo {author} {\bibfnamefont {R.}~\bibnamefont
  {Katilius}},\ }\bibfield  {title} {\bibinfo {title} {Theory of fluctuations
  in nonequilibrium electron gas},\ }\href {https://doi.org/10.1007/BF02724353}
  {\bibfield  {journal} {\bibinfo  {journal} {La Rivista del Nuovo Cimento
  (1978-1999)}\ }\textbf {\bibinfo {volume} {2}},\ \bibinfo {pages} {1}
  (\bibinfo {year} {1979})}\BibitemShut {NoStop}%
\bibitem [{\citenamefont {Kogan}\ and\ \citenamefont
  {Shul’man}(1969)}]{kogan1969theory}%
  \BibitemOpen
  \bibfield  {author} {\bibinfo {author} {\bibfnamefont {S.~M.}\ \bibnamefont
  {Kogan}}\ and\ \bibinfo {author} {\bibfnamefont {A.~Y.}\ \bibnamefont
  {Shul’man}},\ }\bibfield  {title} {\bibinfo {title} {Theory of fluctuations
  in a nonequilibrium electron gas},\ }\href@noop {} {\bibfield  {journal}
  {\bibinfo  {journal} {Sov. Phys. JETP}\ }\textbf {\bibinfo {volume} {29}},\
  \bibinfo {pages} {104} (\bibinfo {year} {1969})}\BibitemShut {NoStop}%
\bibitem [{Note2()}]{Note2}%
  \BibitemOpen
  \bibinfo {note} {In this paper we will use $T$ and $T_0$ interchangeably. The
  latter is preferred when we introduce the local temperature $T(x)$ where
  $T_0$ is defined the boundary values of $T(x)$, and thus the environmental
  temperature $T$.}\BibitemShut {Stop}%
\bibitem [{Note3()}]{Note3}%
  \BibitemOpen
  \bibinfo {note} {In our convention, the distribution function $F$, in the
  equilibrium limit, reduces to $1-2n_F$ where $n_F$ is the usual Fermi-Dirac
  distribution}\BibitemShut {NoStop}%
\bibitem [{\citenamefont {Lifshitz}\ and\ \citenamefont
  {Pitaevskii}(1995)}]{lifshitz1995physical}%
  \BibitemOpen
  \bibfield  {author} {\bibinfo {author} {\bibfnamefont {E.}~\bibnamefont
  {Lifshitz}}\ and\ \bibinfo {author} {\bibfnamefont {L.}~\bibnamefont
  {Pitaevskii}},\ }\href {https://books.google.com/books?id=h7LgAAAAMAAJ}
  {\emph {\bibinfo {title} {Physical Kinetics: Volume 10}}},\ Course of
  theoretical physics\ (\bibinfo  {publisher} {Elsevier Science},\ \bibinfo
  {year} {1995})\BibitemShut {NoStop}%
\bibitem [{\citenamefont {Abrikosov}(2017)}]{abrikosov2017fundamentals}%
  \BibitemOpen
  \bibfield  {author} {\bibinfo {author} {\bibfnamefont {A.}~\bibnamefont
  {Abrikosov}},\ }\href {https://books.google.com/books?id=tTo2DwAAQBAJ} {\emph
  {\bibinfo {title} {Fundamentals of the Theory of Metals}}}\ (\bibinfo
  {publisher} {Dover Publications},\ \bibinfo {year} {2017})\BibitemShut
  {NoStop}%
\bibitem [{\citenamefont {Mahan}(2013)}]{mahan2013many}%
  \BibitemOpen
  \bibfield  {author} {\bibinfo {author} {\bibfnamefont {G.~D.}\ \bibnamefont
  {Mahan}},\ }\href@noop {} {\emph {\bibinfo {title} {Many-particle physics}}}\
  (\bibinfo  {publisher} {Springer Science \& Business Media},\ \bibinfo {year}
  {2013})\BibitemShut {NoStop}%
\bibitem [{\citenamefont {Bruus}\ and\ \citenamefont
  {Flensberg}(2004)}]{bruus2004many}%
  \BibitemOpen
  \bibfield  {author} {\bibinfo {author} {\bibfnamefont {H.}~\bibnamefont
  {Bruus}}\ and\ \bibinfo {author} {\bibfnamefont {K.}~\bibnamefont
  {Flensberg}},\ }\href {https://books.google.com/books?id=v5vhg1tYLC8C} {\emph
  {\bibinfo {title} {Many-Body Quantum Theory in Condensed Matter Physics: An
  Introduction}}},\ Oxford Graduate Texts\ (\bibinfo  {publisher} {OUP
  Oxford},\ \bibinfo {year} {2004})\BibitemShut {NoStop}%
\bibitem [{\citenamefont {Green}\ and\ \citenamefont
  {Das}(2000)}]{Frederick_Green_2000}%
  \BibitemOpen
  \bibfield  {author} {\bibinfo {author} {\bibfnamefont {F.}~\bibnamefont
  {Green}}\ and\ \bibinfo {author} {\bibfnamefont {M.~P.}\ \bibnamefont
  {Das}},\ }\bibfield  {title} {\bibinfo {title} {High-field noise in metallic
  diffusive conductors},\ }\href {https://doi.org/10.1088/0953-8984/12/24/314}
  {\bibfield  {journal} {\bibinfo  {journal} {Journal of Physics: Condensed
  Matter}\ }\textbf {\bibinfo {volume} {12}},\ \bibinfo {pages} {5233}
  (\bibinfo {year} {2000})}\BibitemShut {NoStop}%
\bibitem [{\citenamefont {Sukhorukov}\ and\ \citenamefont
  {Loss}(1999)}]{PhysRevB.59.13054}%
  \BibitemOpen
  \bibfield  {author} {\bibinfo {author} {\bibfnamefont {E.~V.}\ \bibnamefont
  {Sukhorukov}}\ and\ \bibinfo {author} {\bibfnamefont {D.}~\bibnamefont
  {Loss}},\ }\bibfield  {title} {\bibinfo {title} {Noise in multiterminal
  diffusive conductors: Universality, nonlocality, and exchange effects},\
  }\href {https://doi.org/10.1103/PhysRevB.59.13054} {\bibfield  {journal}
  {\bibinfo  {journal} {Phys. Rev. B}\ }\textbf {\bibinfo {volume} {59}},\
  \bibinfo {pages} {13054} (\bibinfo {year} {1999})}\BibitemShut {NoStop}%
\bibitem [{\citenamefont {Schomerus}\ \emph {et~al.}(1999)\citenamefont
  {Schomerus}, \citenamefont {Mishchenko},\ and\ \citenamefont
  {Beenakker}}]{PhysRevB.60.5839}%
  \BibitemOpen
  \bibfield  {author} {\bibinfo {author} {\bibfnamefont {H.}~\bibnamefont
  {Schomerus}}, \bibinfo {author} {\bibfnamefont {E.~G.}\ \bibnamefont
  {Mishchenko}},\ and\ \bibinfo {author} {\bibfnamefont {C.~W.~J.}\
  \bibnamefont {Beenakker}},\ }\bibfield  {title} {\bibinfo {title} {Kinetic
  theory of shot noise in nondegenerate diffusive conductors},\ }\href
  {https://doi.org/10.1103/PhysRevB.60.5839} {\bibfield  {journal} {\bibinfo
  {journal} {Phys. Rev. B}\ }\textbf {\bibinfo {volume} {60}},\ \bibinfo
  {pages} {5839} (\bibinfo {year} {1999})}\BibitemShut {NoStop}%
\bibitem [{\citenamefont {Kamenev}(2023)}]{kamenev2023field}%
  \BibitemOpen
  \bibfield  {author} {\bibinfo {author} {\bibfnamefont {A.}~\bibnamefont
  {Kamenev}},\ }\href@noop {} {\emph {\bibinfo {title} {Field theory of
  non-equilibrium systems}}}\ (\bibinfo  {publisher} {Cambridge University
  Press},\ \bibinfo {year} {2023})\BibitemShut {NoStop}%
\bibitem [{Note4()}]{Note4}%
  \BibitemOpen
  \bibinfo {note} {In the language of field theory, it is identical to the
  second order functional derivative of $\protect \qopname \relax
  o{ln}Z[\protect \bm {A}]$ (as opposed to $Z[\protect \bm {A}]$) with
  respective to the external gauge potential $\protect \bm {A}$, where
  $Z[\protect \bm {A}]$ is the generating function.}\BibitemShut {Stop}%
\bibitem [{\citenamefont {Thielmann}\ \emph {et~al.}(2005)\citenamefont
  {Thielmann}, \citenamefont {Hettler}, \citenamefont {K\"onig},\ and\
  \citenamefont {Sch\"on}}]{PhysRevLett.95.146806}%
  \BibitemOpen
  \bibfield  {author} {\bibinfo {author} {\bibfnamefont {A.}~\bibnamefont
  {Thielmann}}, \bibinfo {author} {\bibfnamefont {M.~H.}\ \bibnamefont
  {Hettler}}, \bibinfo {author} {\bibfnamefont {J.}~\bibnamefont {K\"onig}},\
  and\ \bibinfo {author} {\bibfnamefont {G.}~\bibnamefont {Sch\"on}},\
  }\bibfield  {title} {\bibinfo {title} {Cotunneling current and shot noise in
  quantum dots},\ }\href {https://doi.org/10.1103/PhysRevLett.95.146806}
  {\bibfield  {journal} {\bibinfo  {journal} {Phys. Rev. Lett.}\ }\textbf
  {\bibinfo {volume} {95}},\ \bibinfo {pages} {146806} (\bibinfo {year}
  {2005})}\BibitemShut {NoStop}%
\end{thebibliography}%

\end{document}